\documentclass[final,5p,times,twocolumn]{elsarticle}
\biboptions{sort&compress}

\usepackage{amssymb}
\usepackage{amsmath}
\usepackage{bm}
\usepackage{multirow}
\usepackage{booktabs}
\usepackage{graphicx}
\usepackage{xcolor}
\usepackage{algorithm}
\usepackage{algpseudocode}
\usepackage{array}
\newcommand{\suppheading}[1]{\par\noindent\textit{#1:}\ }
\makeatletter\@ifundefined{bibsep}{}{\setlength{\bibsep}{0pt plus 0.3ex}}\makeatother

\makeatletter
\newcommand{\symauthnote}[1]{\g@addto@macro\@cornotes{%
   \let\thefootnote\relax\footnotetext{#1}}}
\makeatother

\newcommand{\mfm}{mfm}

\journal{Physics Letters B}

\begin{document}

\raggedbottom

\begin{frontmatter}

\title{Directional correlations in nuclear charge-radius model residuals enable extrapolation with calibrated uncertainties}

\author[ism]{D.~Kar$^{\dagger}$}

\author[ism]{S.~Bagchi$^{\ddagger}$}

\author[gsi,jlu]{T.~Dickel}

\author[smu,triumf]{R.~Kanungo}

\symauthnote{$^{\dagger}$\,debodyutikar8@gmail.com}
\symauthnote{$^{\ddagger}$\,sbagchi@iitism.ac.in}

\affiliation[ism]{organization={Department of Physics, Indian Institute of Technology (Indian School of Mines)},
            city={Dhanbad},
            postcode={826004},
            state={Jharkhand},
            country={India}}

\affiliation[gsi]{organization={GSI Helmholtzzentrum für Schwerionenforschung GmbH},
            addressline={Planckstra{\ss}e 1},
            city={Darmstadt},
            postcode={64291},
            state={Hesse},
            country={Germany}}

\affiliation[jlu]{organization={II. Physikalisches Institut, Justus-Liebig-Universit{\"a}t Gießen},
            addressline={Heinrich-Buff-Ring 16},
            city={Gie{\ss}en},
            postcode={35392},
            state={Hesse},
            country={Germany}}

\affiliation[smu]{organization={Department of Astronomy and Physics, Saint Mary’s University},
            city={Halifax},
            postcode={B3H 3C3},
            state={Nova Scotia},
            country={Canada}}

\affiliation[triumf]{organization={TRIUMF},
            city={Vancouver},
            postcode={V6T 2A3},
            state={British Columbia},
            country={Canada}}

\begin{abstract}
We identify a pronounced directional anisotropy in the residuals between
measured nuclear charge radii and two structurally distinct global models: the
phenomenological Weizs\"acker--Skyrme formula (WS*) and the microscopic
Hartree--Fock--Bogoliubov model (HFB-25). In both cases, the residuals remain
correlated over several neutron steps along isotopic chains but decorrelate
almost completely after a single proton step along isotonic chains. This common
pattern, reinforced by a strong correlation between the two residual fields
($r\simeq0.73$), points to a shared deficiency of both global descriptions
rather than a model-specific artefact. Motivated by this geometry, we introduce ARCUS
(Anisotropic Residual Calibration with Uncertainty Scaling), which applies an
anisotropic kernel-regression correction with empirically calibrated prediction
intervals. In out-of-fold cross-validation, ARCUS reduces the root-mean-square
errors of both WS* and HFB-25 by approximately a factor of two. On an
independent, temporally blind test set of 129 nuclei, its prediction intervals
retain coverage close to nominal under extrapolation. An isotropic kernel matched to the same cross-validation coverage instead
substantially overestimates uncertainties on the blind set, showing that reliable
calibration transfer depends on encoding the directional residual structure. In regions with anomalous
structure, such as around $^{52}$Ca, ARCUS keeps its prediction intervals wide
enough to cover the increased errors, rather than yielding overconfident point
predictions. We also extend these
calibrated predictions to 1008 unmeasured nuclei near known isotopic chains,
ranked by uncertainty to support the future
charge-radius measurements.
\end{abstract}


\end{frontmatter}

\section{Introduction}
\enlargethispage{3\baselineskip}
\label{sec:introduction}

The neutron-rich calcium (Ca) isotopes remain one of the clearest examples of how
global charge-radius models can fail in ways that expose missing nuclear-structure
physics. Laser spectroscopy revealed an unexpectedly large increase of
76~\mfm{} in the charge radius from $^{48}$Ca to $^{52}$Ca \cite{Garcia2016}, while
complementary interaction-cross-section measurements later found a rapid growth
of the matter-distribution radii beyond $N=28$ \cite{Tanaka2020}. This makes the
Ca chain a useful reference case for the two global charge-radius
descriptions analysed in this work: the phenomenological Weizs\"acker--Skyrme
charge-radius formula (WS*) \cite{Wang2013,LiWS2025} and the microscopic
Hartree--Fock--Bogoliubov model (HFB-25)
\cite{Goriely2013,Goriely2016,LiHFB2025}. WS* substantially underestimates the
increase from $^{48}$Ca to $^{52}$Ca, whereas HFB-25 gives a prediction closer to the experimental value for
the $^{52}$Ca radius but still leaves structured residuals along the Ca
chain. Often discussed in connection with shell evolution \cite{Otsuka2005}, the Ca anomaly raises a broader question: whether residuals of global charge-radius models contain learnable nuclear-structure information that can improve extrapolations without sacrificing reliable uncertainty estimates near the limits of known isotopic chains.

The nuclear root-mean-square (rms) charge radius
$R_c=\langle r^2\rangle_c^{1/2}$ probes the proton charge distribution and is
sensitive to shell closures, deformation, and pairing
\cite{Angeli2013,Li2021}; modern laser spectroscopy can resolve changes at the
sub-\mfm{} level (\mfm{} $\equiv 10^{-3}$~fm)
\cite{Campbell2016,Neugart2017}. For the charge-radius data analysed here, both models have root-mean-square errors (RMSEs) of order
20--30~\mfm{} and leave visibly structured residuals, consistent with unresolved
structure effects such as shell evolution and shape coexistence
\cite{Otsuka2005,Otsuka2020,Heyde2011}. Machine-learning and related
data-driven methods have improved charge-radius predictions through kernel
ridge regression \cite{Ma2020,Tang2024}, Bayesian neural networks
\cite{Dong2022,Dong2023,Utama2016}, feed-forward neural networks
\cite{Wu2020}, boosted-tree and Gaussian-process regression with symbolic
distillation \cite{Maheshwari2025}, and local isotope-difference relations
\cite{Li2023}, with several of these approaches learning the residuals of
global models. Anisotropic kernels with cross-validated length scales have
been applied to nuclear-mass residuals \cite{WuPan2024} and, more recently,
to charge-radius residuals \cite{Ma2026}. These studies show that global-model
residuals contain exploitable structure, but two methodological gaps remain.
First, in these anisotropic-kernel applications, the directional structure was
inferred within the predictive optimisation rather than independently
characterised beforehand from isotopic and isotonic residual statistics.
Second, charge-radius studies have reported predictive uncertainty bands and
interval-inclusion fractions \cite{Yang2023}, but no study has combined
nominal-versus-observed reliability diagrams and expected calibration error
with a test of whether calibration transfers to a temporally blind set.
Empirical-coverage curves have been used in Bayesian analyses of nuclear
masses \cite{Kejzlar2020} and separation-energy emulators
\cite{Neufcourt2018}, but a comparable calibration-transfer assessment has
not been reported for charge radii.

We first quantify the geometry of the raw experiment--theory residuals using
direction-resolved semivariograms \cite{Cressie1993}, before applying any
machine-learning correction. When viewed as a discrete field over the $(Z,N)$ nuclear chart, the raw residuals show a strong directional anisotropy: they remain correlated over several neutron steps along isotopic chains, but decorrelate almost completely after a single proton step along isotonic chains.

We encode this directional structure in ARCUS (Anisotropic Residual Calibration with Uncertainty Scaling), a calibrated anisotropic-kernel correction to global charge-radius predictions. On the Angeli--Marinova dataset \cite{Angeli2013}, ARCUS reduces the out-of-fold RMSE of both WS* and HFB-25 by approximately a factor of two. Its prediction intervals retain coverage close to nominal on a 129-nucleus temporally blind test set drawn from the compilation of Li et al. \cite{Li2021}, with at most mild over-coverage. In contrast, isotropic kernels calibrated to the same out-of-fold coverage over-cover the blind set far more strongly. The anisotropic intervals thus retain closer-to-nominal coverage without the same loss of sharpness, so directional residual structure supports more reliable calibration beyond nuclei with measured charge radii.

\section{Directional correlations in charge-radius residuals}
\label{sec:anisotropy}

For each nucleus we define the residual
\begin{equation}
\delta(Z,N)=R_{\rm exp}(Z,N)-R_{\rm th}(Z,N).
\label{eq:residual}
\end{equation}
Across the nuclear chart these residuals constitute a spatial random field
\cite{Cressie1993}, which we examine for two complementary global models. The training set contains
885 measured nuclei with $Z\geq 8$ from the Angeli--Marinova compilation
\cite{Angeli2013}, spanning 83 isotopic chains from oxygen (O) to curium (Cm).
The WS* charge-radius formula \cite{Wang2013,LiWS2025} is a
four-parameter phenomenological expression that supplements the leading
$A^{1/3}$ scaling with shell and deformation corrections from the
Weizs\"acker--Skyrme framework; on this data set its RMSE is
21.8~\mfm{}. On the other hand, HFB-25
\cite{Goriely2013,Goriely2016,LiHFB2025} is a self-consistent Hartree--Fock--Bogoliubov model with a Skyrme
energy-density functional, in which the charge radius is obtained from the
computed proton density; its RMSE is 25.4~\mfm{} on the 884 nuclei for which
a tabulated prediction is available ($^{17}$Ne is absent from the HFB-25
table).

We measure the directional structure of the residuals from the raw
experiment--theory field. The empirical semivariogram is defined as \cite{Cressie1993}
\begin{equation}
\gamma_N(h) =
\frac{1}{2}
\left\langle
\left[
\delta(Z,N+h)-\delta(Z,N)
\right]^2
\right\rangle_{\rm same\ isotopic\ chain},
\label{eq:semivar}
\end{equation}
where the average runs over all available pairs of nuclei
separated by $h$ neutron steps (the lag) within the same isotopic chain. The
quantity $\gamma_Z(h)$ is defined analogously for proton-number separations
along isotonic chains. Small $\gamma(h)$ indicates similar residuals at short
separation; as the lag increases, the semivariance rises and levels off at a
height $\sigma_\text{sill}^2$ (the sill). We characterise each direction by fitting the
exponential model $\gamma(h)=\sigma_\text{sill}^2\,(1-e^{-h/\xi})$ and extracting the correlation length
$\xi$\footnote{We use $\xi$ for the correlation length obtained from the
empirical semivariogram and $\ell$ for the internal kernel length-scale
hyperparameters introduced in Sec.~\ref{sec:method}. The two belong to different
functional forms and are not numerically comparable.}.

\begin{figure}[tb]
\centering
\includegraphics[width=\columnwidth]{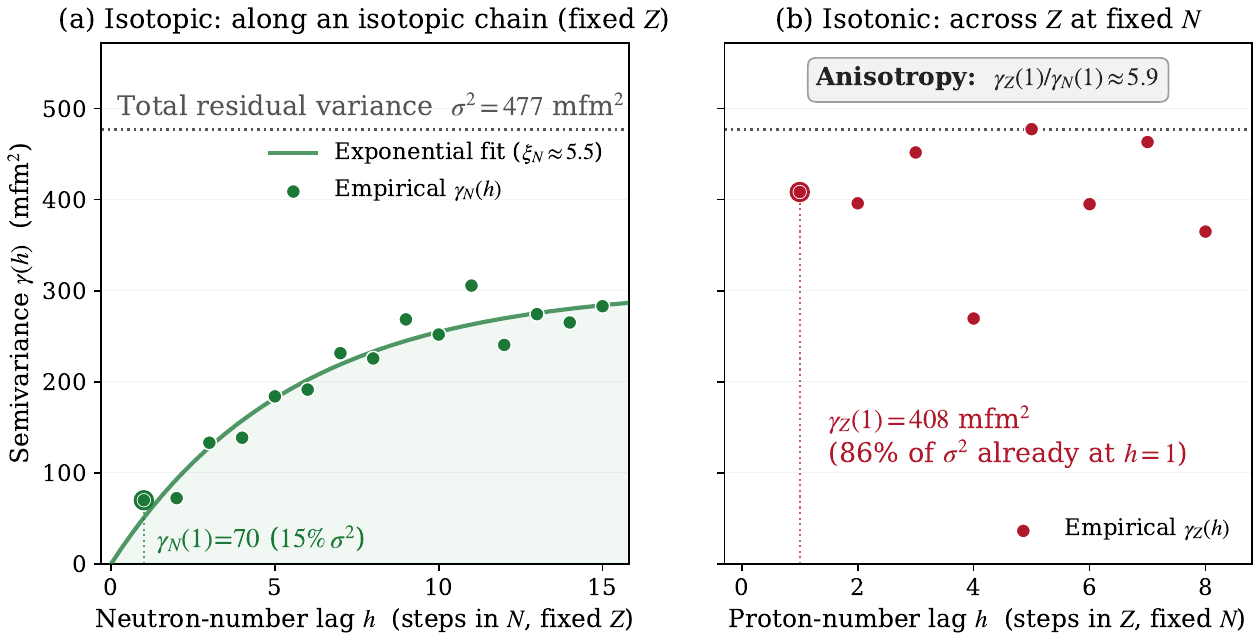}
\caption{
Empirical semivariograms of the raw WS* charge-radius residuals in the isotopic
((a), fixed $Z$) and isotonic ((b), fixed $N$) directions. The isotopic
semivariance rises gradually over a correlation length $\xi_N\simeq5$--6
neutrons, whereas the isotonic semivariance already nears the total residual
variance after a single proton step ($\xi_Z\ll1$). The isotopic semivariance
stays below $\sigma^2$, the signature of zonal anisotropy; HFB-25 shows the same
geometry (Fig.~S3).
}
\label{fig:semivar}
\end{figure}

The two directions behave very differently (Fig.~\ref{fig:semivar}). For
WS*, the total residual variance is $\sigma^2\simeq476~\mfm{}^2$. One neutron
step along an isotopic chain gives $\gamma_N(1)=69.8~\mfm{}^2$, only $15\%$ of $\sigma^2$, with fitted
correlation length $\xi_N\simeq5.5$ neutrons; one proton step across isotonic
chains already gives $\gamma_Z(1)=408~\mfm{}^2$, $86\%$ of $\sigma^2$. A similar
pattern is reproduced by HFB-25, with $\gamma_N(1)=120~\mfm{}^2$,
$\xi_N\simeq6.2$ and $\gamma_Z(1)=484~\mfm{}^2$. In both models, the
proton-direction semivariance reaches its plateau by the first proton step,
consistent with $\xi_Z\ll1$.
These estimates are stable against lag binning and chain selection
(Supplementary Sec.~S2).

The same asymmetry appears directly in quantities that experiments measure. A
model's error on a one-neutron \emph{isotope shift}, the radius change between
adjacent isotopes, is the difference of two residuals, so its RMSE is
$\sqrt{2\gamma_N(1)}$; the one-proton \emph{isotone shift} error is
$\sqrt{2\gamma_Z(1)}$. Both models therefore predict isotope shifts
far better than absolute radii: the one-neutron-shift RMSE is 11.8~\mfm{} for
WS* and 15.5~\mfm{} for HFB-25, against absolute-radius RMSEs of 21.8 and
25.4~\mfm{}. Their one-proton isotone shifts, by contrast, are predicted
\emph{worse} than the absolute radii, at 28.6 and 31.1~\mfm{}. Thus, the residual
correlations retained along isotopic chains directly benefit isotope-shift
predictions. No comparable advantage remains for one-proton isotone shifts, for
which the residuals largely decorrelate.

Equation~(\ref{eq:semivar}) takes differences between residuals along a
fixed-$Z$ chain, so any per-element offset (a systematic model bias shared
by nuclei of the same element) cancels and does not contribute to
$\gamma_N$. This is why the $\gamma_N$ sill lies below $\sigma^2$. The variance
associated with this offset accounts for about
$61\%$ of $\sigma^2$ in both models, despite the larger total variance of
HFB-25 (Supplementary Sec.~S2). A single
proton step changes the element, so this component no longer cancels and
$\gamma_Z(1)$ approaches the total residual variance, a signature of
\emph{zonal anisotropy} \cite{Cressie1993}. Although WS* and
HFB-25 are structurally distinct, both share this offset, and their
residual fields are strongly correlated ($r\simeq0.73$). This agreement suggests
a common source of missing structure in the two global descriptions rather than
an artefact specific to either model.

This anisotropy has a natural structural interpretation, although the extracted
correlation lengths remain empirical. Along isotopic chains, $Z$ is fixed, so
added neutrons modify the charge distribution only indirectly through gradual
changes in shell filling, deformation, and pairing, allowing the residuals to
remain correlated over several neutron steps. Along isotonic chains, each proton
step changes the element and therefore the per-element offset, introducing a
new additive constant rather than a gradual decorrelation. Part of this offset may be
experimental rather than structural, since absolute radii within an
element share chain-wide normalisation systematics \cite{Angeli2013}.

To test whether the loss of residual smoothness reflects rapid structural
evolution, we use tabulated deformations. For each
isotopic chain containing at least four one-neutron steps, 46 chains
in total, we compare the residual roughness (the rms one-neutron-step residual
difference, taken over all adjacent-isotope pairs in the chain) with the mean
local rate of change of the quadrupole deformation $\beta_2$,
$\overline{|\Delta\beta_2/\Delta N|}$, from the finite-range droplet model
FRDM(2012) \cite{Moller2016}. Chains
with more rapidly varying deformation also exhibit rougher residuals, with
Spearman coefficients $\rho=0.65$ for WS* and $0.50$ for HFB-25.
By contrast, the deformation magnitude
$\overline{|\beta_2|}$ is only weakly correlated with residual roughness
($\rho\simeq0.1$; Supplementary Sec.~S2). These results associate residual roughness with the rate of
structural change along a chain rather than with deformation itself: residuals
remain smooth where deformation evolves gradually and become rougher near abrupt
changes.

\section{The ARCUS framework}
\label{sec:method}

ARCUS exploits this directional anisotropy in the nuclear chart to correct the
predicted radii of the theoretical models using the residuals of nearby
measured nuclei. Two offsets
are first removed from the residuals, a parity-class median and a per-element median; the latter
absorbs the per-element offset, which carries most of the residual
variance, and leaves a smaller remainder for the kernel. That remainder stays correlated along an isotopic chain but
largely resets when a proton is added (Sec.~\ref{sec:anisotropy}), so each neighbour
$\mathbf{x}'=(Z',N')$ of a target $\mathbf{x}=(Z,N)$ enters the residual average
with weight $\exp(-d_p)$ (Nadaraya--Watson regression
\cite{Nadaraya1964,Watson1964}), where
\begin{equation}
d_p(\mathbf{x},\mathbf{x}') =
\left[
\left|\frac{\Delta Z}{\ell_Z}\right|^p
+
\left|\frac{\Delta N}{\ell_N}\right|^p
\right]^{1/p}
\label{eq:kernel}
\end{equation}
is the separation between the two nuclei in units of the length
scales $\ell_Z$ and $\ell_N$, with $\Delta Z$ and $\Delta N$ the proton- and
neutron-number differences and $p\ge2$ a shape exponent (elliptical contours at
$p=2$, more box-like for larger $p$). Cross-validation, minimising a continuous
ranked probability score (CRPS)~\cite{Gneiting2007} with a penalty for under-coverage (Supplementary Sec.~S3), returns $\ell_Z\simeq1$ proton and $\ell_N/\ell_Z\simeq10$--$13$
for both WS* and HFB-25, with $p\simeq2.2$--$2.4$; the ratio varies between folds,
but the scale separation $\ell_N\gg\ell_Z$ is robust.

Whereas $d_p$ fixes how strongly each neighbour is weighted, the uncertainty of a prediction
depends on how far the target lies from the measured nuclei. ARCUS records this
as $d_\text{aniso}=\min_i d_p(\mathbf{x}_*,\mathbf{x}_i)$, the distance to the
nearest measured nucleus: small in the interior of the data and growing as predictions
extrapolate beyond it, where the assumption of smooth local correlation is least
supported. Each prediction also carries a raw uncertainty from the scatter of its
neighbours, which is calibrated to the out-of-fold (OOF) errors by isotonic
regression \cite{Niculescu2005}, a monotonic map fitted separately for each
parity class because pairing gives even and odd nuclei different scatter. The
calibrated uncertainty is then widened with $d_\text{aniso}$ and toward the ends
of isotopic chains, to maintain coverage past the training boundary (Supplementary Sec.~S4). Since this calibration is tuned on the OOF data, the OOF
coverage and expected calibration error (ECE) it reaches are consistency checks
rather than a test; the genuine test is the 129-nucleus temporally blind set of
Sec.~\ref{sec:results}. Throughout, $\sigma_{68}$ is the calibrated Gaussian
scale, so the 68\% interval is $\pm z_{68}\sigma_{68}$ with $z_{68}=0.994$,
whereas $\sigma_{95}$ is the 95\% interval half-width and
$\widetilde{\sigma}_{68}$ is the median $\sigma_{68}$ over the evaluated set, a
measure of interval sharpness. Cov$_{68}$/Cov$_{95}$ is the fraction of held-out
nuclei contained, and ECE is the mean absolute gap between observed and nominal
coverage (Supplementary Sec.~S4).

\section{Results}
\label{sec:results}

ARCUS is evaluated using two complementary protocols. The first is five-fold cross-validation on the 885 nuclei from the Angeli--Marinova compilation \cite{Angeli2013}, which provides OOF predictions within the training region. The second is a temporally blind test on 129 nuclei taken from the later compilation of Li et al.\ \cite{Li2021} and not included in Ref.~\cite{Angeli2013}. The blind-test set was fixed before any modelling choices were made. Most of these nuclei lie beyond the previously measured ends of their isotopic chains, so the blind test functions as a controlled extrapolation test of predictive accuracy and uncertainty calibration in the regions probed by radioactive-ion-beam (RIB) facilities.

ARCUS substantially improves the point predictions of both global models. For WS*, the RMSE decreases from 21.8 to 10.3~\mfm{} in OOF validation, while on the blind-test set it decreases from 17.6 to 13.1~\mfm{} (Table~\ref{tab:performance}). A comparable reduction is found for HFB-25, from 25.4 to 13.2~\mfm{} out of fold and from 26.0 to 18.6~\mfm{} on the blind-test set. These results are competitive with published charge-radius methods evaluated on chronological test sets (Sec.~\ref{sec:discussion}).

\begin{figure}[!ht]
\centering
\includegraphics[width=\columnwidth]{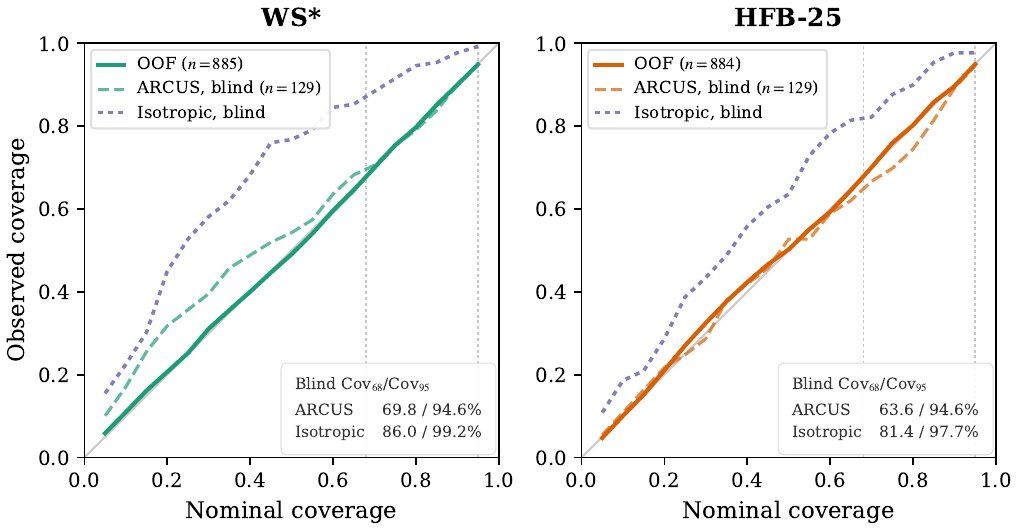}
\caption{Reliability diagrams for WS* (left) and HFB-25 (right) at the reference seed: ARCUS OOF (solid, $n=885/884$), ARCUS blind (dashed, $n=129$), and an isotropic kernel recalibrated to the same OOF coverage, on the same blind set (dotted). Its OOF curve coincides with the ARCUS one and is not drawn. Faint guides mark the 68\% and 95\% levels.}
\label{fig:calib}
\end{figure}

ARCUS additionally provides prediction intervals whose coverage can be tested directly. Out of fold, both baselines are calibrated to nominal, with expected calibration errors of 0.5\% and 0.9\% (medians 0.8\% and 1.2\%). On the blind set, Cov$_{68}$ is 69.8\% for WS* and 63.6\% for HFB-25 (Fig.~\ref{fig:calib}), each within the $\pm4.1$-point binomial standard error at $n\!=\!129$. Across the 45 reruns WS* is mildly conservative (median Cov$_{68}$ 73.6\%) and HFB-25 sits just below nominal (median 66.7\%), with Cov$_{95}$ on nominal for both (median 95.3\%). On the blind set the same diagnostic gives 5.3\% and 1.9\% (medians 9.4\% and 2.5\%). At $n\!=\!129$ sampling alone spreads a perfectly calibrated model's ECE over 1.4--5.5\% (median 2.6\%): the HFB-25 values are indistinguishable from that, and the WS* median lies above it only in the conservative direction, its curve staying at or above nominal at every level.

\begin{table}[tb]
\caption{
Performance summary. Each entry is the value at the reference seed (42),
followed in parentheses by the median over 45 reruns (seeds 1--45).
Improvement factors use the corresponding model RMSE.
Seed ranges and the chain-block bootstrap are in Supplementary Sec.~S6.
}
\label{tab:performance}
\scriptsize
\centering
\setlength{\tabcolsep}{3pt}
\begin{tabular*}{\columnwidth}{@{\extracolsep{\fill}}lrrrrr@{}}
\toprule
Dataset & RMSE & Factor & Cov$_{68}$ & Cov$_{95}$ & $\widetilde{\sigma}_{68}$ \\
        & (\mfm{}) & & (\%) & (\%) & (\mfm{}) \\
\midrule
\multicolumn{6}{c}{WS*} \\
OOF ($n\!=\!885$)   & 10.3 (10.8) & 2.1 (2.0) & 68.0 (67.9) & 94.9 (94.9) & 4.7 (4.7) \\
Blind ($n\!=\!129$) & 13.1 (13.1) & 1.3 (1.3) & 69.8 (73.6) & 94.6 (95.3) & 12.0 (14.4) \\
\midrule
\multicolumn{6}{c}{HFB-25} \\
OOF ($n\!=\!884$)   & 13.2 (13.5) & 1.9 (1.9) & 68.1 (68.1) & 94.9 (94.9) & 6.4 (6.6) \\
Blind ($n\!=\!129$) & 18.6 (18.6) & 1.4 (1.4) & 63.6 (66.7) & 94.6 (95.3) & 12.7 (14.7) \\
\bottomrule
\end{tabular*}
\end{table}

The benefit of anisotropy is clearest in calibration transfer rather than point-prediction accuracy. When constrained to be isotropic and recalibrated on the same OOF residuals, the kernel still achieves nominal OOF coverage. The difference emerges on the temporally blind set: the test RMSE increases only from 13.1 to 14.2~\mfm{} (Table~\ref{tab:ablation}), but the prediction intervals become strongly over-conservative, with Cov$_{68}$/Cov$_{95}$ = 86.0\%/99.2\% on intervals 1.9 times wider (median $\sigma_{68}$ 23.2 against 12.0~\mfm{}; Fig.~\ref{fig:calib}).

This follows from the residual structure of Sec.~\ref{sec:anisotropy}. Residuals vary smoothly along isotopic chains, whereas a single proton step changes the per-element offset and decorrelates them almost completely. An isotropic kernel forces both directions to share one correlation scale, so that offset is read as additional local variability. As same-element support decreases beyond the measured range, this produces increasingly broad intervals. Separating the two directions avoids this, giving the anisotropic kernel near-nominal blind-set coverage with markedly sharper intervals.

\begin{figure}[!ht]
\centering
\includegraphics[width=\columnwidth]{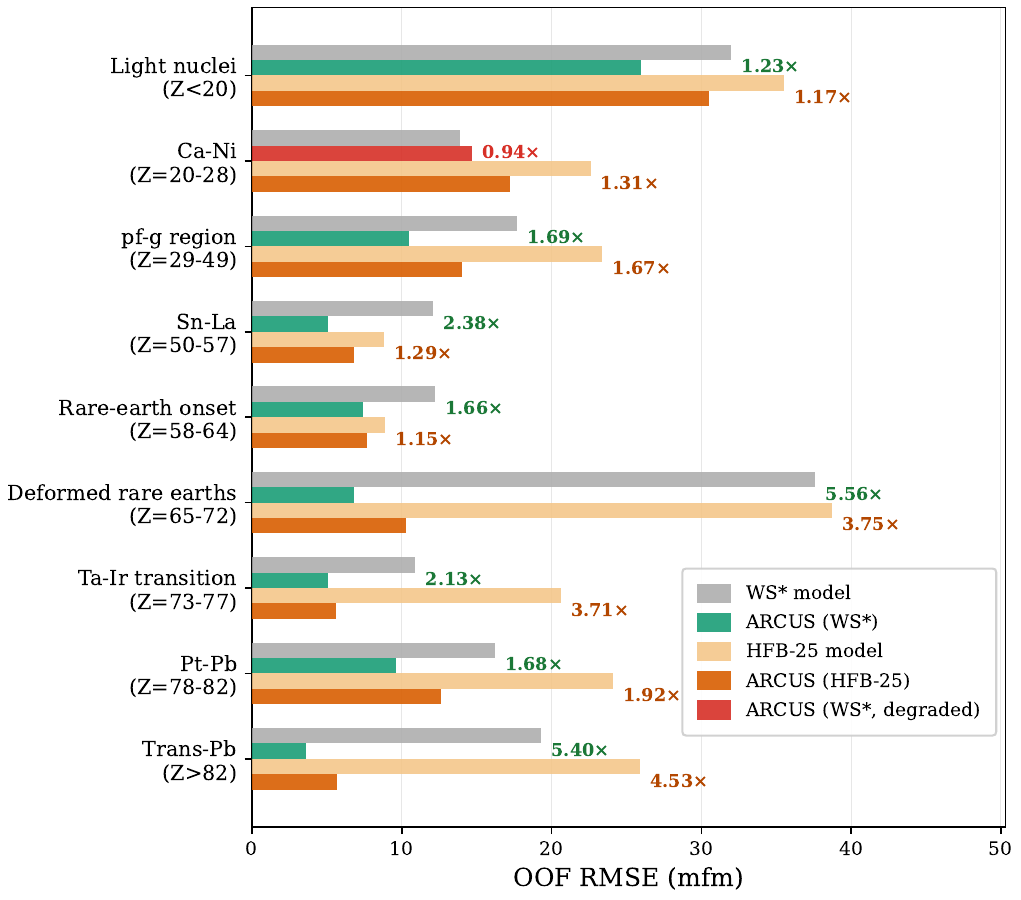}
\caption{Regional OOF improvement factors (baseline RMSE / ARCUS RMSE) for nine non-overlapping nuclear-structure regions. Green/red bars show WS* factors, orange bars HFB-25; the red bar marks the sole WS* degradation (Ca--Ni). ARCUS bars are 45-run medians of the per-seed RMSEs.}
\label{fig:regional}
\end{figure}

\begin{table}[tb]
\caption{Kernel ablation for WS* at the reference seed (OOF $n=885$; test $n=129$). The three configurations differ only in $\ell_N/\ell_Z$ and $p$, sharing the same kernel [Eq.~(\ref{eq:kernel})] and calibration pipeline, and all are matched to OOF Cov$_{68}\approx68\%$; the isotropic recalibration needs a wider search range to reach it (Supplementary Sec.~S4).}
\label{tab:ablation}
\scriptsize
\setlength{\tabcolsep}{3pt}
\begin{tabular*}{\columnwidth}{@{\extracolsep{\fill}}lrrrrr@{}}
\toprule
Configuration & \shortstack[r]{RMSE$_\text{OOF}$\\(\mfm{})} & \shortstack[r]{RMSE$_\text{test}$\\(\mfm{})} & \shortstack[r]{Cov$_{68}^\text{test}$\\(\%)} & \shortstack[r]{Cov$_{95}^\text{test}$\\(\%)} & \shortstack[r]{$\widetilde{\sigma}_{68}^\text{test}$\\(\mfm{})} \\
\midrule
Isotropic ($\ell_N\!=\!\ell_Z,\ p\!=\!2$)        & 10.79 & 14.2 & 86.0 & 99.2 & 23.2 \\
Anisotropic ($\ell_N\!\neq\!\ell_Z,\ p\!=\!2$)   & 10.21 & 13.1 & 71.3 & 94.6 & 12.5 \\
Full ARCUS ($\ell_N\!\neq\!\ell_Z,\ p$ free)     & 10.32 & 13.1 & 69.8 & 94.6 & 12.0 \\
\bottomrule
\end{tabular*}
\end{table}

\begin{figure*}[t]
\centering
\includegraphics[width=\textwidth]{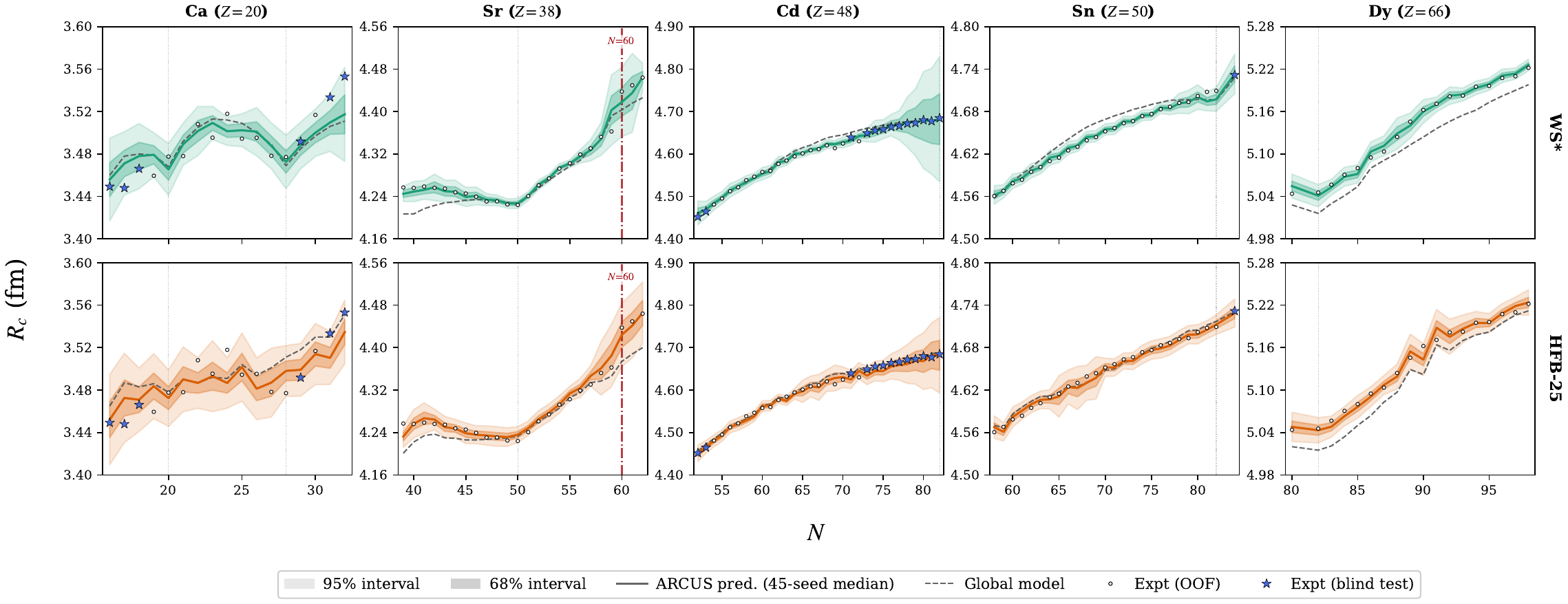}
\caption{Predictions along five isotopic chains for both models. Top row: WS* (teal); bottom row: HFB-25 (orange). Columns: Ca ($Z=20$), Sr ($Z=38$), Cd ($Z=48$), Sn ($Z=50$), Dy ($Z=66$). Solid line = ARCUS prediction; dashed line = global model; shaded bands = 68\%/95\% prediction intervals; open circles = experimental radii (training/OOF nuclei); stars = experimental radii for post-2013 blind-test nuclei. The dash-dotted line in the Sr panel marks the $N=60$ shape transition. The prediction and both bands are 45-run medians.}
\label{fig:chains}
\end{figure*}

The regional analysis shows where smooth residual interpolation is most effective and where its assumptions become less reliable. Figure~\ref{fig:regional} summarises the OOF RMSE reductions across nine non-overlapping nuclear-structure regions, with the full results given in Supplementary Sec.~S5. For WS*, the largest gains occur in the deformed rare-earth ($Z=65$--72) and trans-Pb regions, with improvement factors of approximately 5.4--5.6. HFB-25 shows the same qualitative pattern, consistent with the gradual evolution of collective structure along isotopic chains in these regions. The only regional degradation for WS* occurs in Ca--Ni, with an improvement factor of 0.94, whereas HFB-25 improves by a factor of 1.31. This model dependence marks the same shell-evolution region as the $^{52}$Ca anomaly discussed below. For light nuclei ($Z<20$), cluster, halo, continuum, and isospin-mixing effects may violate the kernel's smoothness assumption; predictions in this region should therefore be regarded as indicative rather than fully calibrated.

\begin{figure}[tb]
\centering
\includegraphics[width=\columnwidth]{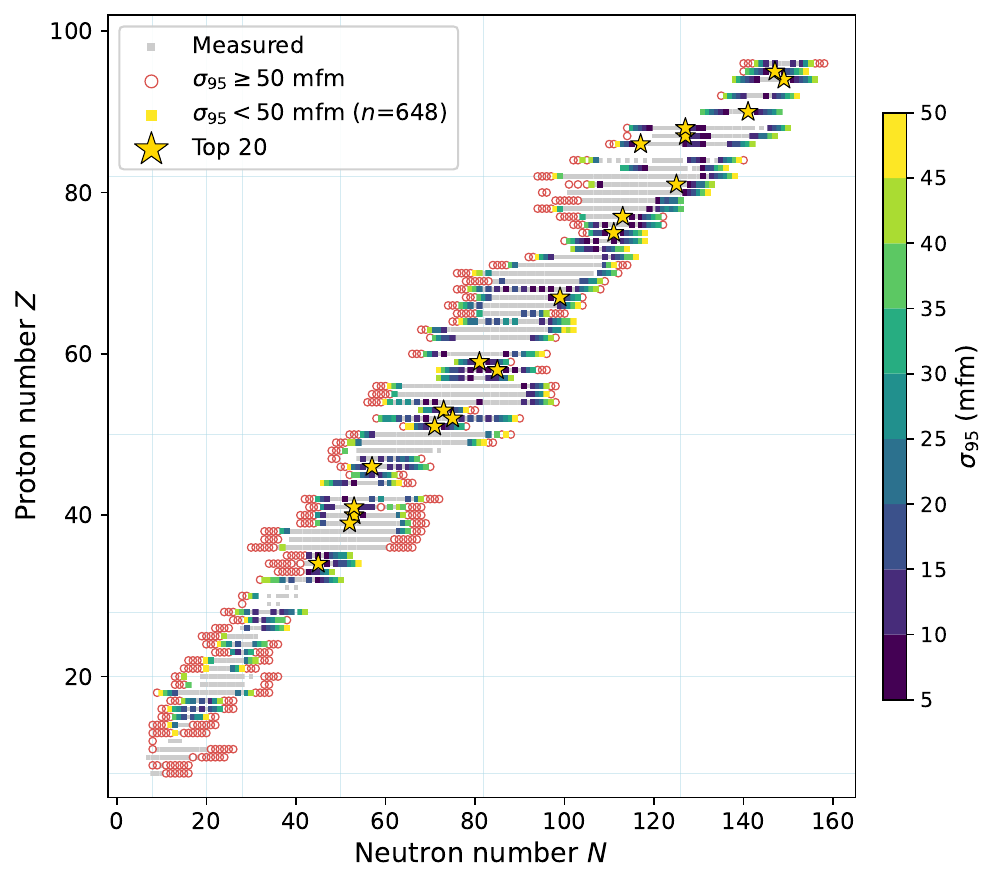}
\caption{Calibrated charge-radius predictions for the 1008 catalogue nuclei defined in the text. Measured nuclei are grey; predictions with $\sigma_{95} \geq 50$~\mfm{} are open red circles; the 648 with $\sigma_{95} < 50$~\mfm{} are colour-coded by $\sigma_{95}$. Gold stars mark the 20 most certain cases. Predictions and intervals are 45-run medians from models trained on the 885 nuclei from the Angeli--Marinova \mbox{compilation~\cite{Angeli2013}}.}
\label{fig:frontier}
\end{figure}

Figure~\ref{fig:chains} compares ARCUS predictions along five isotopic chains. The method reproduces the measured charge-radius trends in structurally distinct systems, ranging from the spherical tin (Sn) chain to the strongly deformed rare-earth dysprosium (Dy) chain. Strontium (Sr) and cadmium (Cd) provide more stringent tests. In Sr, both global models miss the sharp $N=60$ shape transition~\cite{Buchinger1990}, whereas ARCUS reduces the chain RMSE from 24.1 to 10.4~\mfm{} for WS* and from 28.5 to 9.9~\mfm{} for HFB-25. The transition is encoded in the neighbouring measured residuals, so ARCUS reproduces this abrupt structural change within the well-sampled chain. Cd instead tests extrapolation: ten post-2013 isotopes extend beyond the most neutron-rich Cd isotope in the training set and reach the $N=82$ closure. Their blind-test RMSEs are 2.6~\mfm{} for WS* and 9.6~\mfm{} for HFB-25, and all predictions fall within the calibrated 95\% intervals. Calcium is the deliberate counterexample, discussed next.

$^{52}$Ca marks the limit of what local residual interpolation can recover. For WS*, the baseline model predicts a 42-\mfm{} increase in charge radius from $^{48}$Ca to $^{52}$Ca, compared with the measured 76~\mfm{} \cite{Garcia2016}. The pre-2013 training set contains no local evidence of the missing shell-evolution effect, so ARCUS cannot recover it, and its blind prediction differs from experiment by 35.7~\mfm{}. For HFB-25, the situation is reversed: the baseline model predicts the $^{52}$Ca radius to within about 2~\mfm{}, but the smoother trend encoded by neighbouring residuals increases the ARCUS error to 18.4~\mfm{}. Both misses are highly reproducible: across the 45-seed pool the WS* error varies only between 34.3 and 36.9~\mfm{} and the HFB-25 error between 15.9 and 19.7~\mfm{}, so $^{52}$Ca reflects a systematic limitation of local residual interpolation rather than a sampling fluctuation. Neither error, however, is accompanied by an overconfident interval: the standardised residuals are 1.9 for WS* and 1.5 for HFB-25, and both predictions remain inside their calibrated 95\% intervals. Coverage holds here even though nothing in the local data marks $^{52}$Ca as anomalous: it lies only two neutrons beyond the pre-2013 measured range, and its interval is of ordinary width for a blind-test nucleus. Strontium provides the complementary case: its $N=60$ transition is just as abrupt, but it is bracketed by measured neighbours on both sides, and ARCUS recovers it. The limitation at $^{52}$Ca is therefore not the abruptness of the structural change, but the absence of a corresponding local signal in the training data. This is the behaviour that matters for the catalogue predictions provided by ARCUS below, where neither compilation provides a measurement that would reveal a failure of this kind.

The temporally blind test simulates this extrapolation by evaluating nuclei measured only later, and its near-nominal coverage provides empirical support for using the calibrated uncertainties of the nuclei in the catalogue to prioritise future charge-radius measurements. These predictions come from the same five fold models, trained on the Angeli--Marinova compilation \cite{Angeli2013} alone, so the catalogue is generated by the model the blind test validates. The catalogue comprises nuclei that appear in the WS* theoretical table, satisfy $Z\geq8$ and $N\geq8$, and lie within $\lceil \xi_N \rceil = 6$ neutrons of the measured isotopic range for the same element in the Angeli--Marinova compilation. Nuclei with experimental charge radii in either the Angeli--Marinova or Li et al.\ compilation are excluded, so blind-test nuclei inside this window do not enter the catalogue. Radii measured outside these compilations, such as the recent scandium and tin chains \cite{Bai2025,Gustafsson2025}, can therefore overlap with the catalogue. These criteria identify 1008 nuclei, of which 648 (64\%) have $\sigma_{95}<50$~\mfm{}, an uncertainty scale relevant to current collinear laser-spectroscopy measurements at RIB facilities \cite{Campbell2016,Neugart2017}. This count, the catalogue and the ranking below are 45-run medians. The smallest uncertainties, $\sigma_{95}\simeq5$--8~\mfm{}, occur for heavier nuclei adjacent to well-measured isotopic chains. Figure~\ref{fig:frontier} maps the full set, Table~\ref{tab:frontier_top10} lists the twenty most certain predictions, and the complete catalogue is provided as Supplementary Data, described in Sec.~S7.

The catalogue should not be interpreted as a set of equally reliable point predictions. Rather, it provides uncertainty-ranked priorities that reflect the local residual geometry: prediction intervals remain narrow where the residual structure is well constrained and widen where extrapolation is less reliable.

\begin{table}[tb]
\caption{Twenty most certain catalogue predictions (one per element, ranked by $\sigma_{95}$). $R_\text{ARCUS}$ is the calibrated predicted charge radius. Both columns are 45-run medians.}
\label{tab:frontier_top10}
\scriptsize
\begin{tabular*}{\columnwidth}{@{\extracolsep{\fill}}lrr@{\hskip 1.2em}lrr@{}}
\toprule
Nucleus & $R_\text{ARCUS}$ (fm) & $\sigma_{95}$ (\mfm{}) & Nucleus & $R_\text{ARCUS}$ (fm) & $\sigma_{95}$ (\mfm{}) \\
\midrule
$^{206}$Tl & 5.479 &  5.1 & $^{242}$Am & 5.893 &  7.2 \\
$^{93}$Zr  & 4.316 &  6.1 & $^{190}$Ir & 5.396 &  7.2 \\
$^{243}$Pu & 5.883 &  6.8 & $^{126}$I  & 4.745 &  7.3 \\
$^{186}$Re & 5.364 &  6.8 & $^{94}$Nb  & 4.332 &  7.3 \\
$^{214}$Fr & 5.604 &  6.9 & $^{79}$Se  & 4.138 &  7.3 \\
$^{91}$Y   & 4.275 &  6.9 & $^{166}$Ho & 5.210 &  7.3 \\
$^{143}$Ce & 4.914 &  7.0 & $^{127}$Te & 4.728 &  7.4 \\
$^{215}$Ra & 5.613 &  7.0 & $^{203}$Rn & 5.549 &  7.4 \\
$^{231}$Th & 5.771 &  7.0 & $^{140}$Pr & 4.889 &  7.6 \\
$^{122}$Sb & 4.682 &  7.1 & $^{103}$Pd & 4.492 &  7.8 \\
\bottomrule
\end{tabular*}
\end{table}

\section{Discussion and outlook}
\label{sec:discussion}

The regional and chain-by-chain results clarify when residual learning is most effective. ARCUS performs best when the residual structure of the global model is represented in nearby measurements. This includes abrupt changes when they are locally sampled, as for the $N=60$ transition in strontium. When no corresponding signal exists in the training residuals, as for $^{52}$Ca, smooth interpolation cannot reconstruct it. The uncertainty diagnostics should be interpreted accordingly: interval widths reflect the local support available for extrapolation, whereas large standardised residuals identify nuclei for which the smoothness assumption is inadequate. Such limitations arise jointly from the baseline model, the local data coverage, and the kernel assumption; residual learning cannot recover structure absent from both the baseline prediction and nearby measured residuals.

\begin{table}[tb]
\caption{
Protocol-aware comparison of selected machine-learning charge-radius RMSE
values. \textbf{Eval.} denotes the evaluation regime: chrono = chronological
train/test split, OOF-$K$ = out-of-fold predictions under $K$-fold
cross-validation, and LOO = leave-one-out cross-validation, the limiting case
$K=N$. \textbf{Train$\rightarrow$Test} gives the training and evaluation sample
sizes; for OOF-$K$ and LOO, the training and evaluation pools coincide
($n\rightarrow n$). Values from different protocols should
not be read as a strict leaderboard. \textbf{Calib.\ tested} marks papers reporting the
calibration diagnostics compared here: reliability diagrams, expected
calibration error, or empirical coverage at nominal levels. A dash marks
methods that either do not quantify predictive uncertainty or do not assess
its calibration. All rows are
restricted to $Z\ge8$ except those from Ref.~\cite{Maheshwari2025}, whose
evaluation sample retains light nuclei ($Z<8$); their RMSE is therefore not
directly comparable.
}
\label{tab:benchmark}
\scriptsize
\begin{tabular*}{\columnwidth}{@{\extracolsep{\fill}}llllcc@{}}
\toprule
Method & Ref. & Eval. & Train$\rightarrow$Test & RMSE (\mfm{}) & \shortstack[c]{Calib.\\tested} \\
\midrule
EKRR + RCHB          & \cite{Tang2024}       & LOO    & 1014$\rightarrow$1014 & 9.2         & -- \\
ARCUS (WS*)         & this work             & OOF-5  & 885$\rightarrow$885   & 10.3        & \checkmark \\
ARCUS (WS*)         & this work             & chrono & 885$\rightarrow$129   & 13.1        & \checkmark \\
SVGP                 & \cite{Li2025SVGP}     & chrono & 820$\rightarrow$110   & 11.8        & -- \\
LightGBM             & \cite{Li2025SVGP}     & chrono & 820$\rightarrow$110   & 12.5        & -- \\
BNN (D6)             & \cite{Dong2023}       & chrono & 820$\rightarrow$113   & 13.9        & -- \\
CNN                  & \cite{Cao2023CNN}     & chrono & 814$\rightarrow$114   & 15.6        & -- \\
KRR                  & \cite{Ma2020}         & LOO    & 884$\rightarrow$884   & ${\sim}17$  & -- \\
BNN + DFT (ELD)      & \cite{Utama2016}      & chrono & 722$\rightarrow$98    & 26.2        & -- \\
GPR (+symbolic)      & \cite{Maheshwari2025} & OOF-4  & 956$\rightarrow$956   & 31.5$^{a}$  & -- \\
LGBM                 & \cite{Maheshwari2025} & OOF-4  & 956$\rightarrow$956   & 56.1        & -- \\
\bottomrule
\multicolumn{6}{p{0.92\columnwidth}}{$^{a}$\scriptsize Best numerical model from Ref.~\cite{Maheshwari2025}; the paper's distilled symbolic expressions trade accuracy for interpretability, with RMSE ${\sim}70$~\mfm{} for the simplest closed-form expression.}
\end{tabular*}
\end{table}

Table~\ref{tab:benchmark} places the predictive accuracy of ARCUS in the context of selected machine-learning approaches to nuclear charge radii. The reported RMSE values span a wide range, but much of this variation reflects the evaluation protocol. Leave-one-out validation is favourable to local methods because nearly the entire measured chart remains available for each prediction, whereas OOF cross-validation primarily tests interpolation within the sampled region. Chronological splits provide a more direct assessment of extrapolation to newly measured nuclei near the limits of known isotopic chains. ARCUS is competitive under this stricter comparison, but its OOF RMSE should not be interpreted as a direct ranking against results obtained using different protocols. The distinctive comparison instead concerns uncertainty quantification: among the charge-radius studies listed in Table~\ref{tab:benchmark}, ARCUS is the only method whose prediction intervals are assessed using reliability diagrams, expected calibration error, and empirical coverage at nominal levels.

Anisotropic kernels have previously been applied to nuclear-mass residuals \cite{WuPan2024} and, more recently, to charge-radius residuals \cite{Ma2026}, but anisotropy enters ARCUS differently. Separate isotopic and isotonic semivariograms first show that residual correlations persist along isotopic chains but decay rapidly across isotones. This observed geometry provides an empirical basis for distinct neutron- and proton-direction kernel scales, whose numerical values are then selected by cross-validation. The correction remains traceable to the baseline model and nearby measured residuals, while its prediction intervals are calibrated out of fold and tested independently on the temporally blind set. This matters most in extrapolation, where support for any residual correction weakens far from measured nuclei. Rather than masking this reduced support, ARCUS lets its intervals widen with distance from the data while remaining sharper than an isotropic treatment, so the directional residual signal informs both the correction and a faithful account of its uncertainty. ARCUS thus extends recent calibration studies for nuclear masses and separation energies \cite{Neufcourt2018,Kejzlar2020} to charge radii. Measurements reported after the blind set was fixed, including those along the $^{104\text{--}134}$Sn chain \cite{Gustafsson2025} and in neutron-rich scandium \cite{Bai2025}, provide prospective tests of both the predictions and their uncertainties.

\section{Conclusions}
\label{sec:conclusions}

The central result of this work is that the measured directional correlations
in model residuals can improve charge-radius predictions while preserving
uncertainty calibration beyond the training data. ARCUS encodes the structure
common to the phenomenological WS* and microscopic HFB-25 residuals
through an anisotropic-kernel correction with empirically calibrated prediction
intervals. It reduces the OOF RMSE of both models by approximately a factor of
two and improves their predictions on a 129-nucleus temporally blind set. More
importantly, an isotropic kernel can be adjusted to achieve nominal OOF coverage
but substantially over-covers the blind set. ARCUS, by contrast, remains close
to nominal coverage after chronological extrapolation \cite{Angeli2013,Li2021}. In anomalous cases such
as $^{52}$Ca, the uncertainty diagnostics expose the limitations of smooth
residual interpolation rather than concealing them behind an overconfident
prediction.

The resulting catalogue provides calibrated predictions for 1008 nuclei
near known isotopic chains, ranked by uncertainty to support the
prioritisation of future charge-radius measurements. More generally, ARCUS
illustrates a transferable strategy for residual learning: first characterise
the structure of the theory residuals, and then test whether the resulting
uncertainty calibration transfers to data excluded from model development.
The same strategy may apply to other ground-state observables whose model
residuals retain local correlations associated with shell structure, pairing, or
deformation, such as nuclear masses, derived separation energies, and quadrupole
moments, wherever independent validation data exist.

\section*{Data availability}
The public experimental and theoretical data sources used in this work are cited
in the text. The catalogue of ARCUS charge-radius predictions for 1008 nuclei is
provided as Supplementary Data. The ARCUS implementation and trained configurations are
available from the corresponding author upon reasonable request.

\section*{Acknowledgements}

D.~Kar gratefully acknowledges financial support from the Ministry of Education
(MoE), Government of India, through the Institute Assistantship of IIT (ISM)
Dhanbad, and support received through the Sandwich PhD Programme of IIT (ISM)
Dhanbad, the DST--DAAD Project-based Personnel Exchange Programme (PPP), and the
GET\_INvolved Programme of GSI/FAIR, Darmstadt, Germany.

\clearpage
\setcounter{section}{0}
\setcounter{equation}{0}
\setcounter{table}{0}
\setcounter{figure}{0}
\setcounter{algorithm}{0}
\renewcommand{\thesection}{S\arabic{section}}
\renewcommand{\theequation}{S\arabic{equation}}
\renewcommand{\thetable}{S\arabic{table}}
\renewcommand{\thefigure}{S\arabic{figure}}
\renewcommand{\thealgorithm}{S\arabic{algorithm}}
\makeatletter
\renewcommand{\tagform@}[1]{\maketag@@@{\normalfont Eq.~#1}}
\makeatother

\twocolumn[\section*{Supplementary Material}\vspace{1em}]

This Supplementary Material provides the technical information needed to audit the ARCUS workflow. The sections below document the residual-field anisotropy, the hyperparameter and calibration choices, the regional performance breakdown, the statistical robustness checks, and the prediction catalogue.

\section{The ARCUS pipeline}
\label{sm:sec:pipeline}

\begin{figure}[!ht]
\centering
\includegraphics[width=0.85\columnwidth, trim={15 60 5 5}, clip]{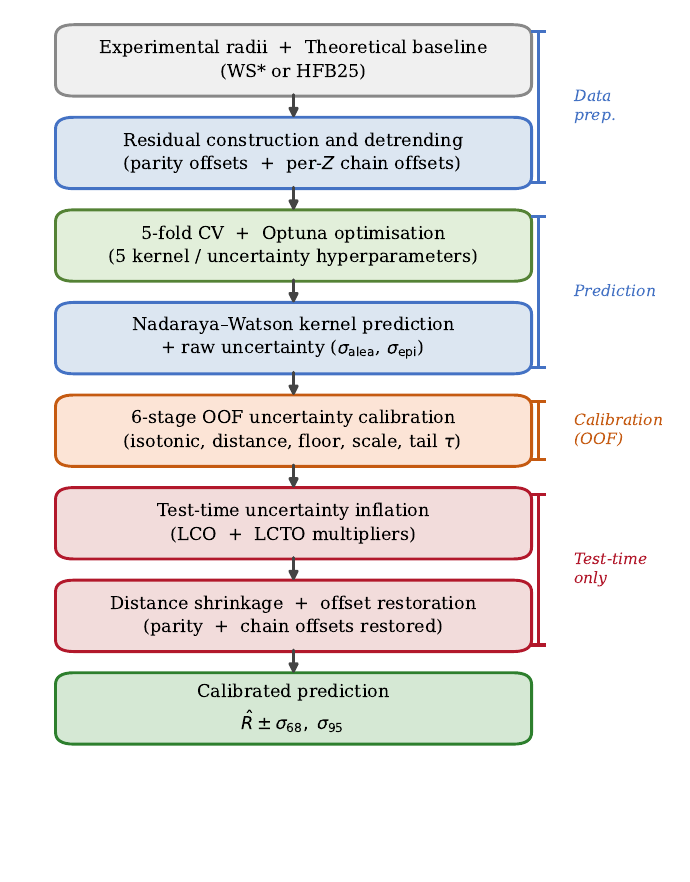}

\caption{ARCUS pipeline. Operations are arranged top-to-bottom in execution
order. The four phases (data preparation, prediction, out-of-fold
uncertainty calibration, test-time inflation) are bracketed at the right. The six-stage
uncertainty calibration is described in detail in Sec.~\ref{sm:sec:uq}; LCO and
LCTO denote the two leave-chain procedures used to fit the test-time
multipliers. Offset restoration is applied to every prediction; distance
shrinkage is active for blind-test and catalogue predictions.}
\label{sm:fig:pipeline}
\end{figure}

ARCUS produces a calibrated residual correction to a global charge-radius baseline. Figure~\ref{sm:fig:pipeline} sketches the pipeline; this section defines the analysis choices needed to understand the workflow. The hyperparameter search and calibration machinery are expanded in Sec.~\ref{sm:sec:hyperopt} and Sec.~\ref{sm:sec:uq}.

\suppheading{Residual field} The training target is the residual of the experimental charge radius against a global theoretical baseline:
\begin{equation}
\delta(Z, N) = R_\text{exp}(Z, N) - R_\text{th}(Z, N),
\label{sm:eq:residual}
\end{equation}
where $R_\text{th}$ is either the phenomenological Weizs\"acker--Skyrme formula
(WS*) or the microscopic Hartree--Fock--Bogoliubov model (HFB-25). The residual
is formed for the 885 nuclei with $Z \geq 8$ from the 2013 compilation~\cite{sm:Angeli2013} (884 for HFB-25, which omits $^{17}$Ne). The 129-nucleus blind test set comprises measurements first reported in the 2021 update~\cite{sm:Li2021}, which were not consulted at any modelling decision.

\suppheading{Detrending} Two systematic offsets are removed before the kernel sees the data. The parity offset
\begin{equation}
\mu_c = \mathrm{median}\{\delta_i : c(Z_i, N_i) = c\},\qquad c \in \{\text{ee, eo, oe, oo}\},
\label{sm:eq:parity_offset}
\end{equation}
absorbs pairing-driven systematics that vary by parity of $Z$ and $N$ but otherwise act as a constant shift within each class. Numerical values are listed in Table~\ref{sm:tab:parity}; both baselines show a clean monotonic ordering from oo (most negative) to ee (most positive), with a smaller spread on HFB-25 because Bogoliubov pairing already absorbs part of the parity dependence self-consistently. After parity removal, a per-element median $\mu_Z$ is subtracted to handle slow chain-level biases (typically larger than $\mu_c$: the per-element offset is the dominant component of the residual variance, Table~\ref{sm:tab:budget}). The kernel then operates on the doubly-detrended residual $\tilde\delta_i = \delta_i - \mu_{c_i} - \mu_{Z_i}$. For out-of-fold (OOF) predictions, both $\mu_c$ and $\mu_Z$ are fitted inside each outer training fold and then applied to the held-out fold, so no held-out residuals enter the detrending. After the OOF loop is complete, the offsets are refit on the full 2013 training set for the blind-test and catalogue-prediction pipelines.

\begin{table}[!ht]
\caption{Parity offsets $\mu_c$ (median residual per class), in mfm. Negative entries indicate the baseline over-predicts the radius for that class.}
\label{sm:tab:parity}
\footnotesize
\centering
\begin{tabular}{l c c}
\toprule
Class               & WS*    & HFB-25  \\
\midrule
ee (even, even)     & $+2.8$ & $+1.7$ \\
eo (even, odd)      & $-1.7$ & $+0.6$ \\
oe (odd, even)      & $-6.1$ & $-2.2$ \\
oo (odd, odd)       & $-8.2$ & $-4.4$ \\
\bottomrule
\end{tabular}
\end{table}

\suppheading{Anisotropic kernel} The kernel weight between two points in the $(Z, N)$ plane is the anisotropic super-elliptic form
\begin{equation}
K(\mathbf{x}, \mathbf{x}') = \exp\bigl[-\min(d_p, d_\text{clip})\bigr],\;\;
d_p = \biggl[\biggl|\frac{\Delta Z}{\ell_Z}\biggr|^p + \biggl|\frac{\Delta N}{\ell_N}\biggr|^p\biggr]^{1/p},
\label{sm:eq:kernel}
\end{equation}
with $\ell_Z, \ell_N$ the directional length scales and $p \geq 2$ the shape exponent. The cap $d_\text{clip} = 2.5$ prevents weights from vanishing in very sparse neighbourhoods. The minimum kernel distance from a query point to any training nucleus, $d_\text{aniso} = \min_i d_p(\mathbf{x}_*, \mathbf{x}_i)$, is recorded as the canonical extrapolation distance and is used by the calibration pipeline (Sec.~\ref{sm:sec:uq}).

\begin{figure}[!ht]
\centering
\includegraphics[width=\columnwidth]{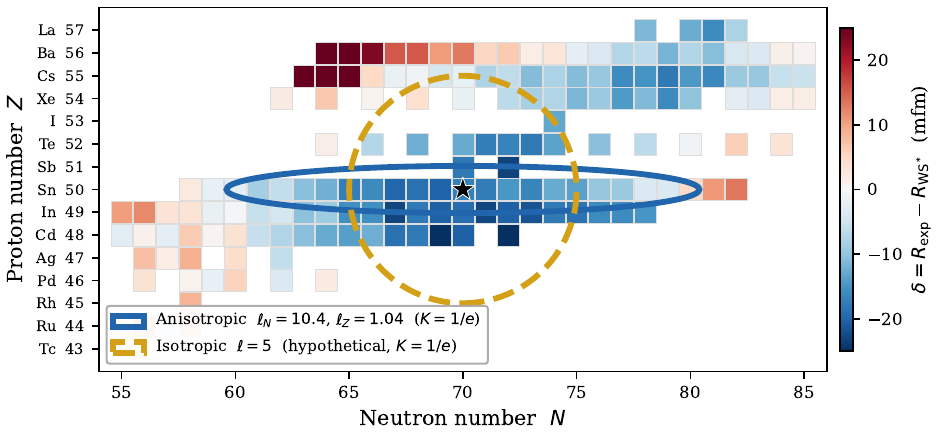}
\caption{Kernel neighbourhood at $^{120}$Sn. Squares: WS* residuals in mfm, plotted only where an experimental charge radius exists; the star marks the $^{120}$Sn query nucleus. Blue solid: anisotropic $K = 1/e$ contour ($\ell_N = 10.4$, $\ell_Z = 1.04$). Gold dashed: a hypothetical isotropic kernel ($\ell_Z = \ell_N = 5$).}
\label{sm:fig:kernel}
\end{figure}

Figure~\ref{sm:fig:kernel} shows the anisotropic neighbourhood centred on $^{120}$Sn. The $K = 1/e$ contour is a narrow super-elliptic neighbourhood extending $\pm 10$ neutrons along the chain while staying within $\pm 1$ proton of the query element. The smooth colour gradient along the chain (WS* residuals vary slowly with $N$) is what the kernel learns from. A hypothetical isotropic kernel ($\ell_N = \ell_Z = 5$, gold dashed) reaches across five elements, mixing residuals with little local correlation in the measured residual field.

\suppheading{Nadaraya--Watson estimator} For a selected neighbour set $\mathcal{S}$, the detrended residual at a query point is the kernel-weighted neighbour average~\cite{sm:Nadaraya1964,sm:Watson1964},
\begin{equation}
\hat\mu(\mathbf{x}_*) = \sum_{i\in\mathcal{S}} w_i\, \tilde\delta_i,\qquad
w_i = (1-\varepsilon)\frac{w_i^\text{raw}}{\sum_{j\in\mathcal{S}} w_j^\text{raw}} + \frac{\varepsilon}{|\mathcal{S}|},\quad i\in\mathcal{S},
\label{sm:eq:nw}
\end{equation}
with $w_i^\text{raw} = K(\mathbf{x}_*, \mathbf{x}_i)$ and a uniform smoothing $\varepsilon = 0.015$ that prevents any single nucleus from dominating. The effective sample size $n_\text{eff} = 1/\sum_{i\in\mathcal{S}} w_i^2$ measures how many independent neighbours actually contribute. The neighbour set is grown adaptively (Algorithm~\ref{sm:alg:neighbor}) until $n_\text{eff} \geq n_\text{eff}^\text{target}$ or a maximum search radius is reached.

Algorithm~\ref{sm:alg:neighbor} gives the analysis-level neighbour search; routine
implementation safeguards, such as empty-set handling, are omitted for clarity.

\begin{algorithm}[!ht]
\caption{ARCUS neighbour expansion and Nadaraya--Watson prediction.}
\label{sm:alg:neighbor}
\begin{algorithmic}[1]
\Require Query $(Z_*,N_*)$; training residuals
$\{(Z_i,N_i,\tilde\delta_i)\}$; kernel parameters
$(\ell_Z,\ell_N,p)$; fixed settings
$n_\text{eff}^\text{target}$, $\varepsilon$, $d_\text{clip}$,
$R_\text{max}$
\Ensure $\hat\mu$, $\sigma_\text{raw}$, $n_\text{eff}$, $d_\text{aniso}$

\State Compute anisotropic distances
$d_i=d_p(\mathbf{x}_*,\mathbf{x}_i)$ for all training nuclei
\State $d_\text{aniso}\gets \min_i d_i$
\For{$R=1$ \textbf{to} $R_\text{max}$}
    \State Select candidates
    $\mathcal{S}\gets\{i:|\Delta Z_i|\le R,\ |\Delta N_i|\le R,\ d_i\le R\}$
    \If{$\mathcal{S}$ is too small}
        \State Store $\mathcal{S}$ if it is the largest candidate set so far
        \State \textbf{continue}
    \EndIf
    \State Compute clipped kernel weights
    $w_i^\text{raw}=\exp[-\min(d_i,d_\text{clip})]$
    \State Normalise and smooth the weights using Eq.~\ref{sm:eq:nw}
    \State $n_\text{eff}\gets 1/\sum_{i\in\mathcal{S}}w_i^2$
    \State \textbf{break} \textbf{if} $n_\text{eff}\ge n_\text{eff}^\text{target}$
\EndFor
\State If the target is not reached, use the largest available candidate set found, recomputing weights if necessary
\State $\hat\mu\gets\sum_{i\in\mathcal{S}}w_i\tilde\delta_i$
\State $\sigma_\text{raw}\gets$ Eq.~\ref{sm:eq:sigma_raw}
\end{algorithmic}
\end{algorithm}

\suppheading{Raw uncertainty} The raw kernel uncertainty decomposes into an aleatoric (within-neighbourhood scatter) and an epistemic (small-$n_\text{eff}$) contribution:
\begin{equation}
\sigma_\text{raw}^2 = \underbrace{\sum_{i\in\mathcal{S}} w_i (\tilde\delta_i - \hat\mu)^2}_{\sigma_\text{alea}^2} + \underbrace{\frac{s_\text{epi}^2}{\max(n_\text{eff}, 1)}}_{\sigma_\text{epi}^2},
\label{sm:eq:sigma_raw}
\end{equation}
with $s_\text{epi}$ a hyperparameter. This $\sigma_\text{raw}$ is then transformed into calibrated $\sigma_{68}$ and $\sigma_{95}$ by the six-stage pipeline of Sec.~\ref{sm:sec:uq}.

\suppheading{Restoration and shrinkage} At prediction time the parity and chain offsets are restored, and the kernel correction is shrunk toward the parity-only baseline according to anisotropic distance:
\begin{equation}
\hat R(\mathbf{x}_*) = R_\text{th}(\mathbf{x}_*) + \mu_{c(\mathbf{x}_*)} + e^{-\gamma d_\text{aniso}}\bigl[\mu_{Z_*} + \hat\mu(\mathbf{x}_*)\bigr],
\label{sm:eq:restore}
\end{equation}
with $\gamma=0$ for OOF predictions and $\gamma=1.0$ for blind-test and
catalogue predictions for both baselines. The latter corresponds to a simple
one-$e$-folding shrinkage in anisotropic kernel distance and was not tuned.
Setting $\gamma=0$ for the blind-test evaluation changes the test root-mean-square error (RMSE) by
$<0.3$\,mfm.
The fixed analysis choices are $\varepsilon=0.015$, $d_\text{clip}=2.5$, $R_\text{max}=15$, five outer folds, three inner folds, 100 Optuna trials, LCTO tail size $k=3$, minimum uncertainty $\sigma_\text{min}=1.5$\,mfm, and the blind-test/catalogue shrinkage rate $\gamma=1.0$ specified above. The free hyperparameters $\ell_Z, \ell_N, p, s_\text{epi}, n_\text{eff}^\text{target}$ are selected per fold; their search bounds and deployed fold summaries are given in Sec.~\ref{sm:sec:hyperopt}.

\section{Directional-correlation diagnostics}
\label{sm:sec:semivar}

The anisotropy of the residual field, claimed in Sec.~2 of the main text, is supported by several complementary analyses: the empirical semivariogram and a variance decomposition of the residual field, the lag-1 autocorrelation, the cross-model residual correlation, and a robustness analysis against analysis choices. They point to the same directional geometry and quantify the per-element offset that the semivariogram cannot display directly.

\suppheading{Semivariogram} Along an isotopic chain the empirical semivariogram~\cite{sm:Cressie1993} is
\begin{equation}
\gamma_N(h) = \frac{1}{2|\mathcal{P}_N(h)|}\sum_{(i,j)\in\mathcal{P}_N(h)}\bigl[\delta(Z_i,N_i) - \delta(Z_j, N_j)\bigr]^2,
\label{sm:eq:semivar}
\end{equation}
where $\mathcal{P}_N(h) = \{(i,j) : Z_i = Z_j,\, N_j - N_i = h\}$ collects all pairs at lag $h$ within the same chain; this is the explicit pair-sum form of the compact ensemble average given as Eq.~(2) of the main text. The isotonic semivariogram $\gamma_Z(h)$ is defined analogously across chains. These are fit-free, pre-ARCUS diagnostics: they require no smoothing or choice of kernel basis, and they are computed on raw theory residuals before any ARCUS step. They motivate the anisotropic kernel geometry used in the Letter; they are not fitted by ARCUS.

Tabulated values for both baselines are in Table~\ref{sm:tab:semi}. At lag $h = 1$ the WS* residuals along chains satisfy $\gamma_N(1) = 70$\,\mfm$^2$, only 15\,\% of the total variance $\sigma^2 = 476$\,\mfm$^2$; across chains, $\gamma_Z(1) = 408$\,\mfm$^2$, already 86\,\%. The HFB-25 numbers reproduce this normalised pattern (19\,\% / 75\,\%) on a different baseline whose total residual variance is larger.

\begin{table}[!ht]
\caption{Empirical semivariograms of raw residuals, in mfm$^2$. Total residual variance: $\sigma^2_\text{WS*} = 476$\,\mfm$^2$, $\sigma^2_\text{HFB-25} = 643$\,\mfm$^2$. The $Z$-direction sample is exhausted at $h = 8$ because the chart spans only a limited number of isotones.}
\label{sm:tab:semi}
\footnotesize
\centering
\begin{tabular}{r r r r r}
\toprule
& \multicolumn{2}{c}{WS*} & \multicolumn{2}{c}{HFB-25} \\
\cmidrule(lr){2-3}\cmidrule(lr){4-5}
$h$ & $\gamma_N$ & $\gamma_Z$ & $\gamma_N$ & $\gamma_Z$ \\
\midrule
1   & 70  & 408 & 120 & 484 \\
2   & 72  & 396 & 122 & 513 \\
3   & 133 & 452 & 197 & 572 \\
5   & 184 & 477 & 237 & 499 \\
8   & 226 & 365 & 262 & 403 \\
12  & 240 & --- & 370 & --- \\
15  & 283 & --- & 379 & --- \\
\bottomrule
\end{tabular}
\end{table}

\suppheading{Exponential fit} An exponential model $\gamma(h) = \sigma_\text{sill}^2[1 - \exp(-h/\xi)]$, in which
the semivariance rises over a correlation length $\xi$ towards a plateau (the
sill) of height $\sigma_\text{sill}^2$, fits the isotopic direction with $\xi_N = 5.5$ neutrons (WS*) and $\xi_N = 6.2$ (HFB-25); the isotonic direction yields $\xi_Z \ll 1$ for both baselines. The isotopic semivariance stays below the total residual variance (\textit{zonal anisotropy}), while the isotonic semivariance approaches $\sigma^2$ already at the first lag, consistent with near-complete decorrelation across proton number. The fitted isotopic sill ($\sigma_\text{sill}^2 \approx 0.64\sigma^2$ for WS*, $0.66\sigma^2$ for HFB-25) should not be read as the within-chain variance, because residuals drift slowly along each chain (quantified below); we obtain the per-element offset directly, by variance decomposition.

\begin{figure}[!ht]
\centering
\includegraphics[width=\columnwidth]{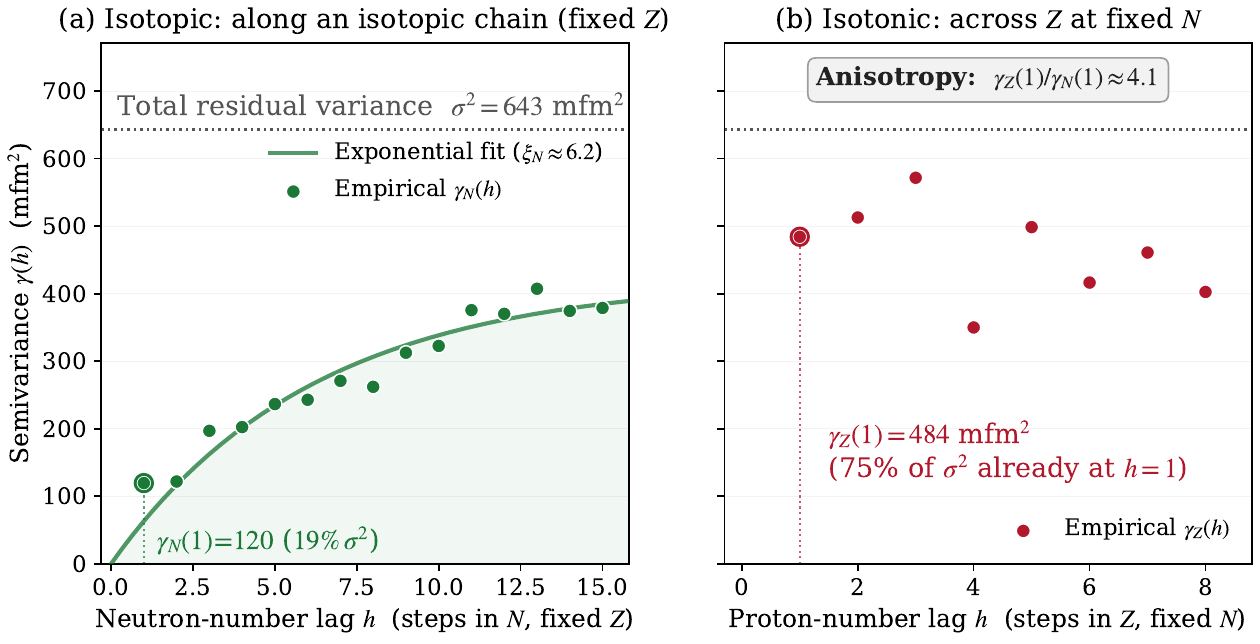}
\caption{Empirical semivariograms of the raw \emph{HFB-25} residuals in the isotopic ((a), fixed $Z$) and isotonic ((b), fixed $N$) directions, the model-independence companion to main-text Fig.~1 (WS*). The zonal-anisotropy pattern replicates on this fully-microscopic baseline: $\gamma_N$ rises over $\xi_N \simeq 6.2$ neutrons and stays below $\sigma^2$, whereas $\gamma_Z$ reaches ${\sim}75\%$ of $\sigma^2$ after a single proton step ($\xi_Z \ll 1$). Plotted values are those of Table~\ref{sm:tab:semi}.}
\label{sm:fig:semivar_hfb25}
\end{figure}

\suppheading{Variance partition} Because a per-element offset is constant along an isotopic chain, it cancels in Eq.~\ref{sm:eq:semivar} and does not appear in $\gamma_N$ at any lag. Its size is obtained directly, by partitioning the total residual variance $\sigma^2$ (via the law of total variance) into a between-element component, the variance of the per-element mean residual, and a within-element remainder. The between-element offset accounts for $61\%$ of $\sigma^2$ for both baselines (Table~\ref{sm:tab:budget}): it is the dominant term, and the same fraction despite the larger total variance of HFB-25. The within-element remainder divides further into a slow drift of the residual along each chain (a non-zero per-chain slope in $N$) and a shorter-range fluctuation that decorrelates over $\xi_N$; the two are comparable in size. Per-chain linear fits find a drift beyond twice its standard error in roughly half of the chains with at least five isotopes. This drift is what makes $\gamma_N$ keep rising rather than saturate, inflating the fitted sill so that it exceeds the true within-element variance ($\approx0.39\,\sigma^2$); the naive estimate $\sigma^2 - \sigma_\text{sill}^2$ would therefore understate the offset. Tentatively, the dominant per-element offset is consistent with an element-specific (proton-structure) systematic that smooth global models do not resolve, while the within-chain drift reflects the neutron-number evolution of nuclear shape (it tracks $|\Delta\beta_2/\Delta N|$, the step-to-step change in the quadrupole
deformation parameter $\beta_2$), not the isospin $\langle N-Z\rangle$ dependence; we treat these as interpretation rather than established attributions.

\begin{table}[!ht]
\caption{Variance budget of the raw residual field, from the law-of-total-variance partition. Percentages are of the total residual variance $\sigma^2$ ($476$\,\mfm$^2$ for WS*, $643$\,\mfm$^2$ for HFB-25).}
\label{sm:tab:budget}
\footnotesize
\centering
\begin{tabular}{l r r r r}
\toprule
& \multicolumn{2}{c}{WS*} & \multicolumn{2}{c}{HFB-25} \\
\cmidrule(lr){2-3}\cmidrule(lr){4-5}
Component & \mfm$^2$ & \%\,$\sigma^2$ & \mfm$^2$ & \%\,$\sigma^2$ \\
\midrule
Per-element offset (between-chain) & 291 & 61 & 393 & 61 \\
Within-chain drift in $N$          &  89 & 19 & 105 & 16 \\
Within-chain fluctuation           &  96 & 20 & 145 & 23 \\
\midrule
Total $\sigma^2$                   & 476 & 100 & 643 & 100 \\
\bottomrule
\end{tabular}
\end{table}

\suppheading{Model independence} The two baselines are constructed from independent frameworks, but their residual fields are not statistically independent. Over the common nuclei the WS* and HFB-25 residuals correlate at $r = 0.73$ (Pearson; the per-element offsets alone at $0.77$). A common-component model $\delta = c + u$, with a model-specific term $u$, attributes a shared variance of ${\approx}\,400$\,\mfm$^2$ to $c$: $85\%$ of the WS* residual variance and $62\%$ of the HFB-25 variance. The replication of the semivariogram geometry across the two baselines is therefore evidence that the geometry is not an artefact of a single theoretical framework, but it is not two statistically independent detections of the same signal: the near-equal offset fractions reflect a residual field that the independently-constructed models largely share. Consistent with the main text, that shared field is itself the finding, a deficiency common to both frameworks, rather than a coincidence to be discounted. This shared structure is not merely an artefact of the common experimental radii: the difference field $\delta_\text{WS*} - \delta_\text{HFB-25} = R_\text{HFB-25} - R_\text{WS*}$ removes $R_\text{exp}$ entirely yet still carries a $30\%$ between-element component, so the two theories genuinely diverge in an element-structured way. We cannot, however, separate within $c$ the physics both models miss from any element-dependent structure in the experimental compilation itself.

\suppheading{Lag-1 autocorrelation} A complementary nearest-neighbour diagnostic is the Pearson correlation between residuals at $(Z, N)$ and the immediate-neighbour pair, separately along each direction (Table~\ref{sm:tab:autocorr}). For WS*, the isotopic value $r_N = 0.85$ corresponds to $r_N^2 = 73$\,\% linear shared variance between neighbouring residuals along a chain; the isotonic value $r_Z = 0.16$ gives only $r_Z^2 = 2.5$\,\%. Under the independent-pair approximation, a Fisher-$z$ comparison gives $z = 19.1$ for WS* and $z = 15.0$ for HFB-25, confirming that the directional difference is far larger than the sampling uncertainty. The relation $\gamma_N(1) \approx \sigma^2(1 - r_N) = 70$\,\mfm$^2$ closes the loop with the semivariogram entry of Table~\ref{sm:tab:semi}.

\begin{table}[!ht]
\caption{Lag-1 Pearson autocorrelation of raw residuals; $n$~=~consecutive-neighbour pair count.}
\label{sm:tab:autocorr}
\footnotesize
\centering
\begin{tabular}{l c c c c c c}
\toprule
Baseline & $r_N$ & $n_N$ & $r_Z$ & $n_Z$ & $r_N/r_Z$ & Fisher~$z$ \\
\midrule
WS*   & 0.85 & 683 & 0.16 & 541 & 5.4 & 19.1 \\
HFB-25 & 0.81 & 682 & 0.25 & 541 & 3.3 & 15.0 \\
\bottomrule
\end{tabular}
\end{table}

\suppheading{Robustness} The fitted isotopic scale $\xi_N$ is stable against the analysis choices behind Eq.~\ref{sm:eq:semivar}, at the level relevant for the main-text claim. Single-lag binning gives $\xi_N = 5.48$ (WS*) and $6.16$ (HFB-25); adjacent-lag binning that halves the number of fit points gives $5.8$ and $6.4$. Restricting the pair pool to chains with at least five measured isotopes (excluding 14 of the 73 chains with two or more isotopes, ${\sim}0.6$\,\% of pairs) gives $5.5$ and $6.2$. A chain-block bootstrap~\cite{sm:Efron1979,sm:Kunsch1989} that resamples entire isotopic chains and refits the exponential model (1000 replicates, seed = 42) gives median $\xi_N = 5.8$ with interquartile range $[4.5,\,9.1]$ for WS* and median $5.7$ with interquartile range $[3.9,\,9.6]$ for HFB-25. The bootstrap distribution is right-skewed, and we do not quote a two-sided 95\,\% interval: in replicates that underweight the longest chains, $\gamma_N$ need not level off within the sampled lags, and the sill--range trade-off of the exponential model is then poorly constrained. Throughout these checks the proton-direction scale remains below one proton step; in the bootstrap it stays below 0.8 protons in all 1000 replicates for both baselines. The qualitative residual geometry (a multi-neutron isotopic scale and near-step isotonic decorrelation) is invariant under these choices.

\suppheading{Deformation-gradient correlation} The main text tests the
smooth-evolution interpretation by correlating chain-level residual roughness
with the local deformation gradient. For every isotopic chain, roughness is
the rms of the one-neutron-step residual differences
$\delta(Z,N{+}1)-\delta(Z,N)$ over consecutive pairs, and the deformation
gradient is the mean of $|\beta_2(Z,N{+}1)-\beta_2(Z,N)|$ over the same
pairs, with $\beta_2$ taken from the finite-range droplet model FRDM(2012)~\cite{sm:Moller2016}. Chains with at
least four one-neutron steps (adjacent-isotope pairs) are retained. The Spearman rank
correlation between roughness and gradient is stable against this retention
threshold: requiring $\geq 4$, $\geq 5$, and $\geq 6$ such steps
retains 46, 41, and 40 chains and gives $\rho = 0.65$, $0.61$, $0.61$
(all $p \leq 4\times10^{-5}$) for WS* and $\rho = 0.50$, $0.40$, $0.42$
(all $p \leq 1.1\times10^{-2}$) for HFB-25. The mean deformation magnitude
$\overline{|\beta_2|}$ of a chain is never significantly correlated with its
roughness ($|\rho| \leq 0.26$, $p \geq 0.1$ across both models and all three
thresholds). The association is therefore specific to the \emph{rate of
change} of deformation, as expected if within-chain residual correlation
reflects smooth structural evolution, and is not a proxy for the presence of
static deformation. Because the FRDM(2012) $\beta_2$ values are themselves
model outputs, we read this as a consistency test of the interpretation
rather than a mechanism determination.

\section{Hyperparameter optimisation}
\label{sm:sec:hyperopt}

The five free hyperparameters of the kernel and the raw-uncertainty model are optimised per outer fold using Optuna~\cite{sm:Optuna2019} with a Tree-structured Parzen Estimator (TPE)~\cite{sm:Bergstra2011} sampler. The search bounds are listed in Table~\ref{sm:tab:search}. Each Optuna trial trains ARCUS on the inner training set, using three-fold inner cross-validation within each outer fold, and is scored by an objective that combines central accuracy, sharpness, and under-coverage control.

\begin{table}[!ht]
\caption{Hyperparameter search space.}
\label{sm:tab:search}
\footnotesize
\centering
\begin{tabular}{l l l l}
\toprule
Parameter            & Symbol                       & Range                            & Scale  \\
\midrule
$Z$-length scale     & $\ell_Z$                     & $[0.8,\, 1.5]$                   & linear \\
Anisotropy ratio     & $\ell_N/\ell_Z$              & $[1.0,\, 15.0]$                  & linear \\
Shape exponent       & $p$                          & $[2.0,\, 3.5]$                   & linear \\
Epistemic scale      & $s_\text{epi}$               & $[3{\times}10^{-4},\, 0.016]$    & log    \\
Target $n_\text{eff}$ & $n_\text{eff}^\text{target}$ & $[5.0,\, 6.0]$                   & linear \\
\bottomrule
\end{tabular}
\end{table}

\suppheading{Objective} The continuous ranked probability score (CRPS) for a
Gaussian predictive distribution at observation $y$ has the closed form $\text{CRPS}(\mu, \sigma, y) = \sigma[z(2\Phi(z)-1) + 2\varphi(z) - 1/\sqrt{\pi}]$, with $z = (y-\mu)/\sigma$. CRPS is a strictly proper scoring rule~\cite{sm:Gneiting2007} that jointly evaluates accuracy and sharpness. The fold objective adds a penalty against under-coverage on the inner-validation predictions:
\begin{equation}
L = \overline{\text{CRPS}} + 0.25\bigl[\max(0,\, 0.68-\hat c_{68})^2 + \tfrac{1}{2}\max(0,\, 0.95-\hat c_{95})^2\bigr],
\label{sm:eq:objective}
\end{equation}
where $\hat c_{68}, \hat c_{95}$ are observed coverages on the inner-validation set. Only under-coverage is penalised explicitly in the inner objective; over-broad raw uncertainties are disfavoured through the CRPS sharpness term. The 0.25 prefactor balances CRPS sharpness against under-coverage without allowing the calibration term to dominate the search. Since the residual semivariogram already points to a strongly anisotropic geometry, moderate changes to this weight are not expected to change the qualitative conclusion that $\ell_N \gg \ell_Z$.

\suppheading{Fold-wise optima} Table~\ref{sm:tab:hyperparam_summary} summarises the deployed hyperparameters as fold medians with fold ranges. The anisotropy ratio $\ell_N/\ell_Z$ varies across folds because once $\ell_Z \lesssim 1$, neighbouring proton chains are already effectively decoupled and the objective is comparatively flat in $\ell_N/\ell_Z$. The robust conclusion is therefore not a precise numerical ratio, but the separation of scales $\ell_N \gg \ell_Z$ for both baselines.

\begin{table*}[t]
\caption{Summary of fold-wise hyperparameter optima for the reference ARCUS configuration (seed = 42). Values are medians across the five outer folds; parentheses give the fold range. The wide span in $\ell_N/\ell_Z$ reflects a flat objective once neighbouring proton chains are effectively decoupled.}
\label{sm:tab:hyperparam_summary}
\scriptsize
\centering
\setlength{\tabcolsep}{4pt}
\begin{tabular}{l c c c c c}
\toprule
Baseline & $\ell_Z$ & $\ell_N/\ell_Z$ & $p$ & $s_\mathrm{epi}$ & $n_\mathrm{eff}^{\mathrm{target}}$ \\
\midrule
WS*   & 1.04 (0.83--1.16) & 10.0 (3.1--14.7)  & 2.20 (2.06--3.15) & 0.0100 (0.0075--0.0112) & 5.06 (5.04--5.35) \\
HFB-25 & 0.93 (0.84--1.03) & 12.7 (5.2--13.8) & 2.36 (2.00--2.82) & 0.0157 (0.0137--0.0160) & 5.09 (5.04--5.46) \\
\bottomrule
\end{tabular}
\end{table*}

\section{Uncertainty calibration}
\label{sm:sec:uq}

The raw kernel uncertainty $\sigma_\text{raw}$ from Eq.~\ref{sm:eq:sigma_raw}
is heteroscedastic but not yet calibrated against the actual prediction errors.
This section details the uncertainty-calibration component of ARCUS, which maps
$\sigma_\text{raw}$ to calibrated $\sigma_{68}$ and $\sigma_{95}$ using
out-of-fold (OOF) prediction errors. The final blind-test and catalogue uncertainties
then apply additional inflation to account for queries that lie beyond the
training boundary.

\suppheading{Stages 1--2, parity-stratified isotonic regression} The OOF
predictions yield pairs $(\sigma_{\text{raw},i},\, |e_i|)$, where $e_i$ is
the OOF prediction error. We fit a stacked isotonic regression~\cite{sm:Niculescu2005}
within each parity class $c$, using $\log\sigma_\text{raw}$ as the feature and
$|e|$ as the target, and convert the predicted mean absolute error to a
standard deviation under the half-normal relation:
\begin{equation}
\sigma_\text{cal} = \mathrm{ISO}_c(\log\sigma_\text{raw})\cdot \sqrt{\pi/2},
\label{sm:eq:iso}
\end{equation}
with floor $\sigma_\text{cal} \geq 1.5$\,mfm. Class stratification is used
because pairing gives the four parity classes different residual \emph{scatter};
the class offsets of Table~\ref{sm:tab:parity} are already removed by detrending, so
it is the class-dependent spread, not the offset, that motivates stratifying here.
A global isotonic fit would not preserve these class-dependent uncertainty scales.

\suppheading{Stages 3--4, distance scaling and additive floor} A multiplicative distance term $\sigma_\text{cal}\to\sigma_\text{cal}(1 + \alpha_\text{dist} d_\text{aniso}^2)$ and an additive floor
$\sigma^2 \to \sigma^2 + c_d d_\text{aniso}^2 + c_n/n_\text{eff}$
allow the calibrated uncertainty to grow with extrapolation distance and neighbour sparsity. The distance coefficient $\alpha_\text{dist}$ is tuned first to flatten 68\,\% coverage across five $d_\text{aniso}$ bins. The additive-floor coefficients $c_d$ and $c_n$ are then chosen by grid search against the overall and high-distance 68\,\% coverage, subject to minimum floor terms that prevent vanishing extrapolation uncertainty. For both baselines the search returns zero, so the floors set the deployed values. WS* selects $\alpha_\text{dist}=0$, so the additive floor carries the distance dependence; HFB-25 selects $\alpha_\text{dist}=1.2$.

\suppheading{Stage 5, global scale} A single scalar $g_\text{level}$ is found
by binary search so that the OOF 68\,\% coverage matches the nominal level.
The default search bracket is $[0.5, 2.5]$. If the search reaches an endpoint
of that bracket, it is widened to $[0.1, 5.0]$ so that the configuration is
still recalibrated to nominal OOF coverage. This is required for the isotropic
ablation of main-text Table~2, where widening is what keeps it a test of
blind-test calibration transfer rather than a test of whether OOF coverage can
be fitted. The deployed anisotropic pipelines reach nominal OOF coverage
within the default bracket at the reference seed, with $g_\text{level}$ between
0.83 and 0.89 for the two baselines. The output of this stage is
$\sigma_{68}=g_\text{level}\sigma_\text{floored}$, where $\sigma_\text{floored}$ denotes the
uncertainty after Stages 1--4 (isotonic calibration, distance scaling and
additive floor).

\begin{figure}[!ht]
\centering
\includegraphics[width=\columnwidth]{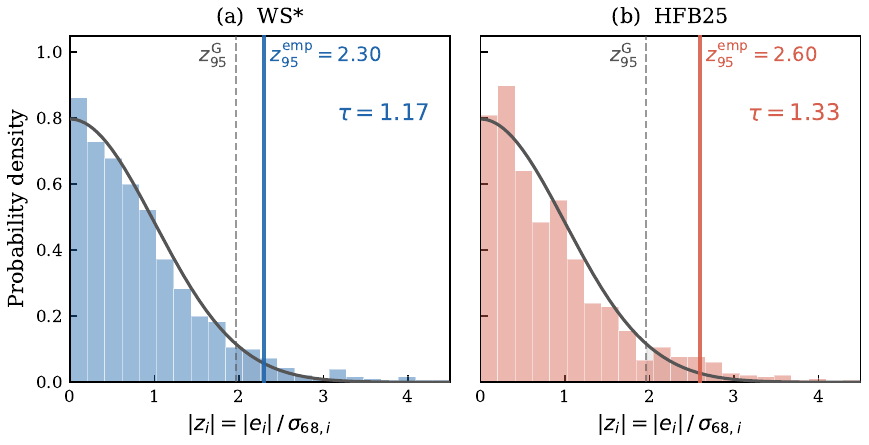}
\caption{Standardised OOF residual $|z_i| = |e_i|/\sigma_{68,i}$ for
(a)~WS* and (b)~HFB-25. Solid curve: half-normal reference ($\tau=1$).
Dashed grey: Gaussian 95th percentile $z_{95}=1.96$; coloured solid:
empirical 95th percentile. The ratio of the empirical to Gaussian 95th
percentile defines $\tau$.}
\label{sm:fig:tail}
\end{figure}

\suppheading{Stage 6, tail factor} The standardised residual
$|z_i| = |e_i|/\sigma_{68,i}$ may have heavier tails than a Gaussian
reference. We account for this with a tail factor
\begin{equation}
\tau = \max\bigl(1,\; Q_{95}(|z|)/z_{95}\bigr),\qquad
\sigma_{95} = \sigma_{68}\,\tau\,z_{95},
\label{sm:eq:tau}
\end{equation}
where $Q_{95}(|z|)$ is the empirical 95th percentile of the OOF
standardised residuals and $z_{95} = \Phi^{-1}(0.975) = 1.96$. Note the
convention used throughout: $\sigma_{68}$ is the calibrated Gaussian
\emph{scale}, so that the 68\,\% interval is $\pm z_{68}\sigma_{68}$ with
$z_{68} = \Phi^{-1}(0.84) = 0.99$ and standardised residuals are
$e_i/\sigma_{68,i}$, whereas $\sigma_{95}$ above is already an interval
half-width. Both baselines require mild
non-Gaussian tail inflation in the deployed run, with $\tau = 1.17$ for
WS* and $\tau = 1.33$ for HFB-25. Figure~\ref{sm:fig:tail} shows the standardised
OOF residual distributions and the empirical tail excess used to define $\tau$.

\suppheading{Test-time inflation with LCO and LCTO} OOF residuals diagnose interpolation, whereas blind-test and catalogue predictions often lie at unseen chain edges. Two leave-chain procedures estimate the additional inflation needed in this extrapolative regime. \textit{Leave-chain-out} (LCO) holds out an entire isotopic chain at a time, fits ARCUS and refits the full $\sigma$-calibration stack on the remaining chains, and predicts the held-out chain; the aggregated standardised residuals define a chain-scale inflation $g_\text{lco}$ and characteristic distance $d_\text{lco}$. \textit{Leave-chain-tail-out} (LCTO) holds out the $k = 3$ most neutron-rich nuclei of each chain with $\geq 9$ measured isotopes, and separately the $k = 3$ most neutron-deficient; the aggregated residuals as a function of $d_\text{beyond}$, the neutron distance beyond the retained chain boundary, define a chain-edge inflation $g_\text{edge}$. For blind-test and catalogue nuclei the same symbol denotes the distance beyond the full measured 2013 boundary of the chain. The applied multiplier is $m_\text{test} = \max(m_\text{lco}, m_\text{edge})$, where each multiplier grows linearly from 1 at the training boundary to its fitted value at the characteristic distance. The final test-time uncertainty is
$\sigma_{68}^\text{test} = g_\text{level}\,\sigma_\text{floored}\,m_\text{test}$,
where $\sigma_\text{floored}$ includes the isotonic calibration, distance scaling, and additive floor from Stages 1--4. The corresponding $\sigma_{95}^\text{test}$ follows from Eq.~\ref{sm:eq:tau}. Because the LCO and LCTO holdouts are deliberately more demanding than ordinary OOF interpolation, $g_\text{lco}$ and $g_\text{edge}$ are conservative estimates of the inflation required at test time rather than universal extrapolation constants, and blind-test and catalogue intervals at large $d_\text{aniso}$ are correspondingly cautious. All calibration parameters are listed in Table~\ref{sm:tab:calib}.

\begin{table}[!ht]
\caption{Calibration parameters for the reference run (seed = 42). These parameters define the reported intervals and are not interpreted as independent physics observables.}
\label{sm:tab:calib}
\footnotesize
\centering
\begin{tabular}{l c c}
\toprule
Parameter             & WS*               & HFB-25             \\
\midrule
$g_\text{level}$      & 0.893             & 0.826             \\
$\tau$ (tail factor)  & 1.173             & 1.325             \\
$\alpha_\text{dist}$  & 0.0               & 1.2               \\
$c_d$ (fm$^2$)        & $1.6\times10^{-5}$ & $8.4\times10^{-5}$ \\
$c_n$ (fm$^2$)        & $1.2\times10^{-5}$ & $2.7\times10^{-5}$ \\
$g_\text{lco}$        & 2.125             & 1.000             \\
$d_\text{lco}$        & 0.97              & 1.07              \\
$g_\text{edge}$       & 1.533             & 1.359             \\
$d_\text{lcto}$       & 2.0               & 2.0               \\
\bottomrule
\end{tabular}
\end{table}

\suppheading{Expected calibration error (ECE)} ECE is the mean absolute deviation between observed and nominal coverage at $M = 15$ levels equally spaced on $[0.10, 0.95]$:
\begin{equation}
\text{ECE} = \frac{1}{M}\sum_{m=1}^{M}\bigl|\hat c(p_m) - p_m\bigr|.
\label{sm:eq:ece}
\end{equation}
For levels $p_m \leq 0.68$, coverage is evaluated using $\sigma_{68}$ directly. For $p_m > 0.68$, the scale is multiplied by an interpolated tail factor $\tau^{t_m}$, where $t_m$ increases linearly from 0 at 68\,\% coverage to 1 at 95\,\% coverage. This makes the ECE diagnostic consistent with the non-Gaussian tail correction in Eq.~\ref{sm:eq:tau}. The OOF ECE values reported in the main text (0.5\,\% for WS*, 0.9\,\% for HFB-25) are computed by Eq.~\ref{sm:eq:ece} with this tail-corrected coverage curve.

\section{Regional performance breakdown}
\label{sm:sec:perf}

\suppheading{Regional improvement} Tables~\ref{sm:tab:regional_ws} and~\ref{sm:tab:regional_hfb25} give the numerical regional breakdown behind Fig.~3 and the discussion in Sec.~4 of the main text. The pattern supports the main-text interpretation: ARCUS gains most where collective structure evolves smoothly along isotopic chains, while regions with rapid structural change or light-nucleus effects require more cautious interpretation.
 
\begin{table}[!ht]
\caption{Regional OOF performance, WS* baseline. Factor $=$ baseline RMSE\,/\,ARCUS RMSE; values $<1$ indicate degradation. RMSE in mfm; $n$~=~training nuclei in region. The ARCUS column is the median of the per-seed RMSEs over the 45-seed pool, so the factor is the median factor; the baseline column does not depend on the partition.}
\label{sm:tab:regional_ws}
\footnotesize
\centering
\begin{tabular}{l c c c c c}
\toprule
Region              & $Z$ range & $n$  & WS*  & ARCUS         & Factor       \\
\midrule
Light nuclei        & $<20$     & 64   & 32.0 & 26.0          & 1.23         \\
Ca--Ni              & 20--28    & 45   & 13.9 & 14.7 & 0.94 \\
pf--g region        & 29--49    & 208  & 17.7 & 10.5          & 1.69         \\
Sn--La              & 50--57    & 126  & 12.1 & 5.1           & 2.38         \\
Rare-earth onset    & 58--64    & 75   & 12.2 & 7.4           & 1.66         \\
Deformed rare earths & 65--72    & 129  & 37.6 & 6.8  & 5.56 \\
Ta--Ir transition   & 73--77    & 25   & 10.9 & 5.1           & 2.13         \\
Pt--Pb              & 78--82    & 114  & 16.2 & 9.6           & 1.68         \\
Trans-Pb            & $>82$     & 99   & 19.3 & 3.6  & 5.40 \\
\bottomrule
\end{tabular}
\end{table}

\begin{table}[!ht]
\caption{Regional OOF performance, HFB-25 baseline. Factor $=$ baseline RMSE\,/\,ARCUS RMSE; RMSE in mfm; $n$~=~training nuclei in region. As in Table~\ref{sm:tab:regional_ws}, the ARCUS column is the 45-seed median of the per-seed RMSEs.}
\label{sm:tab:regional_hfb25}
\footnotesize
\centering
\begin{tabular}{l c c c c}
\toprule
Region ($Z$ range)         & $n$ & HFB-25 & ARCUS         & Factor \\
\midrule
Light ($<20$)              & 63  & 35.5  & 30.5          & 1.17   \\
Ca--Ni (20--28)          & 45  & 22.6  & 17.2          & 1.31   \\
pf--g region (29--49)          & 208 & 23.4  & 14.0          & 1.67   \\
Sn--La (50--57)         & 126 & 8.8   & 6.8           & 1.29   \\
Rare-earth onset (58--64)          & 75  & 8.9   & 7.7           & 1.15   \\
Deformed rare earths (65--72)        & 129 & 38.7  & 10.3 & 3.75 \\
Ta--Ir transition (73--77)  & 25  & 20.6  & 5.6           & 3.71   \\
Pt--Pb (78--82)         & 114 & 24.1  & 12.6          & 1.92   \\
Trans-Pb ($>82$)          & 99  & 25.9  & 5.7  & 4.53 \\
\midrule
Overall                    & 884 & 25.4  & 13.5          & 1.88   \\
\bottomrule
\end{tabular}
\end{table}

The single regional degradation, Ca--Ni on WS* at factor 0.94, is baseline-dependent: the same region improves by 1.31$\times$ on HFB-25. This is physically informative rather than merely negative. It indicates that the ARCUS smooth-kernel correction reaches its limit when the local residual field contains shell-evolution structure that is not smooth in the available training data, and that this limitation depends on what the baseline has already captured. The asymmetry across baselines is therefore itself a useful diagnostic.

\section{Robustness checks}
\label{sm:sec:robust}

\suppheading{Chain-block statistical robustness} Because neighbouring nuclei
within an isotopic chain are correlated, the primary robustness estimate
resamples entire isotopic chains rather than individual nuclei. This
chain-block bootstrap directly tests whether the OOF error reduction survives
the along-chain autocorrelation documented in Table~\ref{sm:tab:autocorr}. The
improvement-factor lower bounds remain well above unity, 1.64$\times$ for
WS* and 1.53$\times$ for HFB-25, so the roughly two-fold OOF RMSE reduction
survives chain-level resampling (Table~\ref{sm:tab:significance}). As a
consistency check, the same chain-block resampling on the 129-nucleus blind
test gives RMSE-improvement intervals of 1.35$\times$ [1.11,\,1.64] for WS*
and 1.40$\times$ [1.15,\,1.63] for HFB-25, so the central-error improvement
also survives chain-level resampling on truly held-out post-2013 data. Thus
the central-error reduction does not rely on treating all nuclei as
independent. Nucleus-level paired tests give much smaller formal $p$-values,
but those assume independent nuclei and are inflated by the strong lag-1
isotopic autocorrelation; the chain-block bootstrap is therefore the primary
robustness check.

\begin{table}[!ht]
\caption{Statistical robustness of the OOF error reduction under a chain-block
bootstrap. Entire isotopic chains are resampled as blocks, preserving the
dominant along-chain correlation measured in Table~\ref{sm:tab:autocorr}.
Confidence intervals (CI) use 10000 chain-block resamples with seed = 42. RMSE and
mean absolute error (MAE) are in mfm. Improvement is the baseline RMSE divided by the ARCUS RMSE.
The chain-block bootstrap $p$-value is the fraction of resamples for which the
mean paired absolute-error improvement,
$|e_i^\text{baseline}|-|e_i^\text{ARCUS}|$, is non-positive. Cohen's $d$ is
computed from per-chain mean absolute-error improvements.}
\label{sm:tab:significance}
\footnotesize
\centering
\begin{tabular}{l r r}
\toprule
Statistic / Metric         & WS*                 & HFB-25              \\
\midrule
RMSE [95\,\% CI]           & 10.3 [8.4,\,12.3]   & 13.2 [10.9,\,15.7] \\
MAE [95\,\% CI]            & 6.1 [5.2,\,7.2]     & 7.9 [6.8,\,9.3]    \\
Improvement                & $2.1\times$ [1.64,\,2.76] & $1.9\times$ [1.53,\,2.42] \\
Chain-block bootstrap ($p$) & $<10^{-4}$          & $<10^{-4}$         \\
Cohen's $d$ (chain-mean)   & 0.57                & 0.62               \\
\bottomrule
\end{tabular}
\end{table}

The chain-block intervals are wider than the nucleus-level bootstrap intervals,
as expected from the strong isotopic autocorrelation, and correspond to an
effective sample size of roughly 40\,\% of the nominal nucleus count. The
central-error reduction holds at that reduced effective sample size. The
temporally blind test in the Letter
remains the primary check of extrapolative performance and calibration
transfer.

\suppheading{Seed robustness} As a stress check, the full stochastic pipeline
(hyperparameter search, fold
assignment, and every calibration stage) was rerun over 45 independent seeds
for both baselines and for the isotropic ablation
(Table~\ref{sm:tab:seedrobust}). The seed set is 1--45 and contains the reference
seed used throughout the Letter, so the single-run values quoted there and the
distributions quoted here describe one sample rather than two. Central
predictions are stable: blind-test RMSE
varies by less than 0.4\,mfm across seeds for the deployed configurations, and by
less than 0.7\,mfm for the isotropic ablation. Coverage is stable on the scale
that a test set of this size can resolve. With $n=129$ blind-test nuclei the
binomial standard error on a nominal 68\,\% rate is
$\sqrt{0.68\cdot0.32/129}\approx4\,\%$, so a perfectly calibrated model measured
on a different draw of 129 nuclei would itself move over a $\pm2\sigma$ window
${\sim}16\,\%$ wide. The observed seed
ranges, 67.4--79.1\,\% for WS* (median 73.6\,\%) and 59.7--73.6\,\% for
HFB-25 (median 66.7\,\%), span $11.6$ and $14.0$ points, both
\emph{narrower} than that window: rerunning the entire stochastic pipeline moves
Cov$_{68}$ less than the irreducible sampling uncertainty of any single
129-nucleus measurement. Both baselines sit close to nominal, on opposite sides of
it: WS* is mildly conservative and falls below 68\,\% at one seed of 45,
while HFB-25 sits marginally below, at 86 of 129 nuclei against a
nominal 88, and falls below at 30 seeds of 45. Both offsets are small next to
the $\pm4$-point standard error of any single 129-nucleus test, and Cov$_{95}$
is close to nominal for both (median 95.3\,\%, range 90.7--97.7\,\%).
The contrast with
the isotropic ablation is categorical for both baselines: isotropic
Cov$_{68}$ is 82.2--91.5\,\% (WS*) and 78.3--87.6\,\% (HFB-25), neither range
overlapping the anisotropic Cov$_{68}$ at any of the 45 seeds, with a median
Cov$_{95}$ of 100.0\,\%
(97.7--100.0\,\% across seeds) against 95.3\,\% for both
corresponding anisotropic kernels.
The intervals themselves differ in width: the median blind $\sigma_{68}$ is
14.4\,mfm for WS* against 22.9\,mfm for its isotropic ablation (11.2--18.8
versus 19.6--31.1 across seeds), and 14.7 against 22.1\,mfm for HFB-25
(11.8--17.4 versus 19.8--28.9), so the isotropic intervals are wider by factors
of $1.6$ and $1.5$ at the median with no overlap at any seed.
Anisotropy therefore yields intervals that are both sharper and better
calibrated, not merely more conservative ones, and the calibration-transfer
conclusion holds at every one of the 45 seeds.

\suppheading{Reference seed and released products} Any single partition is one
draw from this distribution, and the reference seed
sits on its favourable side: its blind
$\sigma_{68}$ and Cov$_{68}$ lie near the tenth percentile for both baselines,
and its WS* blind ECE is the lowest of the 45. That seed was fixed before any of
these reruns were performed and has not been changed since. Rather than report
the released products from that single run, we build them from the pool median:
the regional and chain figures, the catalogue map,
the top-twenty table and the released catalogue. The median is well defined for
per-nucleus predictions and interval widths and, for the ranking, measurably
more representative than any single partition (Sec.~S7). For the regional
figure the median is taken over the per-seed RMSEs rather than over the
predictions, because the plotted quantity is a ratio of RMSEs and the RMSE of a
median prediction would sit below a typical single run. The reference seed is
retained where a median would be ill-defined or misleading: Tables~1 and 2 and
the reliability diagram report it directly, because pointwise medians of a
reliability curve cancel sampling noise and yield a curve closer to the diagonal
than any real run.
Seed-to-seed OOF-coverage variation, fitted by construction,
is not treated as independent evidence.

\suppheading{Coverage versus extrapolation distance} Blind-test coverage is not
uniform in extrapolation distance, so the aggregate
coverages quoted above are marginal, that is, averages over that gradient. Both
kernels show the same gradient, which identifies it as a property of the
distance recalibration and not of anisotropy. Per-bin coverage carries a binomial standard
error of 7--12 percentage points at these sample sizes and is correspondingly
noisy; interval width separates the two kernels more cleanly.
Table~\ref{sm:tab:sharpness_distance} gives the median $\sigma_{68}$ over the same
bins. The isotropic ablation is wider in every bin and for both baselines, by
factors of 1.7--2.8, and by 2.24 (WS*) and 1.59 (HFB-25) over the 113 blind
nuclei beyond the pre-2013 measured range. The anisotropic kernel also shows $\sigma_{68}$
growing with extrapolation distance, from 4.2 to 17.9 mfm for WS* and from 6.0
to 28.4 mfm for HFB-25, as intended by the LCO and LCTO stages.

\begin{table}[!ht]
\centering
\small
\caption{Seed robustness of the blind-test results: median over 45 complete
pipeline reruns (seeds 1--45, which include the reference seed); the
seed-to-seed ranges are quoted in the text above. Each baseline is
shown with its ARCUS (anisotropic, reference) kernel, identical to the
Full ARCUS configuration of main-text Table~2, and its isotropic (wide-bracket
recalibrated) ablation on adjacent rows, so the effect of removing anisotropy
is directly visible. RMSE and median $\sigma_{68}$ in mfm; coverage in \%. The $\sigma_{68}$ entry is the median over the 129 blind nuclei within a run, then the median across runs. Central predictions are
seed-stable. Under seed stress the isotropic kernel is uniformly worse for both
baselines: it raises the blind-test RMSE, inflates the coverage
(over-covering, wider-than-nominal intervals), and widens the intervals themselves
by a factor of about $1.5$--$1.6$; neither its Cov$_{68}$ range nor its median
$\sigma_{68}$ range overlaps the corresponding anisotropic range at any of the
45 seeds. Per-seed values are
provided with the reproduction outputs.}
\label{sm:tab:seedrobust}
\scriptsize
\setlength{\tabcolsep}{4pt}
\begin{tabular}{ll ccc}
\toprule
Baseline & Kernel & RMSE & Cov$_{68}$ & Median $\sigma_{68}$ \\
\midrule
WS*    & ARCUS       & 13.1 & 73.6 & 14.4 \\
       & Isotropic   & 14.5 & 86.8 & 22.9 \\
\midrule
HFB-25 & ARCUS       & 18.6 & 66.7 & 14.7 \\
       & Isotropic   & 21.1 & 82.9 & 22.1 \\
\bottomrule
\end{tabular}
\end{table}

\begin{table}[!ht]
\centering
\caption{Interval sharpness versus extrapolation distance for the reference run.
$d_\text{beyond}$ is the number of neutrons beyond the measured boundary of the
same isotopic chain; $n$ is the number of the 129 blind nuclei in each bin
(identical for both baselines). Entries are the median $\sigma_{68}$ in mfm over
the nuclei in the bin, for the ARCUS (anisotropic) kernel and for the
equivalently calibrated isotropic ablation, with their ratio. Blind-test
coverage for the same configurations is reported in
Table~\ref{sm:tab:seedrobust}: the isotropic ablation over-covers in every bin
beyond the pre-2013 measured range and in both baselines, so its wider intervals do not
buy better calibration.}
\label{sm:tab:sharpness_distance}
\scriptsize
\setlength{\tabcolsep}{5pt}
\begin{tabular}{lc ccc ccc}
\toprule
 & & \multicolumn{3}{c}{WS*} & \multicolumn{3}{c}{HFB-25} \\
\cmidrule(lr){3-5}\cmidrule(lr){6-8}
$d_\text{beyond}$ & $n$ & ARCUS & Isotropic & Ratio & ARCUS & Isotropic & Ratio \\
\midrule
0              &  16 &   4.2 &  10.9 & 2.59 &   6.0 &  14.4 & 2.39 \\
1--2           &  43 &  10.6 &  18.9 & 1.79 &   9.2 &  15.5 & 1.69 \\
3              &  15 &  13.1 &  23.2 & 1.77 &  10.5 &  18.5 & 1.75 \\
4--5           &  23 &  13.1 &  22.1 & 1.69 &  14.8 &  26.5 & 1.78 \\
$\geq 6$       &  32 &  17.9 &  49.3 & 2.75 &  28.4 &  47.4 & 1.67 \\
\midrule
all $\geq 1$   & 113 &  13.0 &  29.1 & 2.24 &  16.0 &  25.5 & 1.59 \\
\bottomrule
\end{tabular}
\end{table}

\section{Charge-radius prediction catalogue}
\label{sm:sec:frontier}

The ARCUS prediction catalogue accompanies the Letter as Supplementary Data: a CSV file
of calibrated charge-radius predictions for 1008 nuclei. The catalogue comprises nuclei that
(i) appear in the WS* theoretical table, (ii) have no measurement in the
2013~\cite{sm:Angeli2013} or 2021~\cite{sm:Li2021} compilations, (iii) satisfy
$Z \geq 8$ and $N \geq 8$, and (iv) lie within $\lceil\xi_N\rceil = 6$ neutrons of the
measured 2013 chain boundary in the same element. The neutron-distance
criterion uses the rounded empirical isotopic semivariogram scale as a locality
cutoff. It is a conservative rule for restricting the catalogue to nuclei close
to measured chains, not a universal correlation-length boundary.

Each catalogue nucleus receives an ensemble prediction from the five fold
models, each trained on the full 885-nucleus dataset with its own fold's
hyperparameters,
\begin{equation}
\hat R = K_\text{fold}^{-1}\sum_{k=1}^{K_\text{fold}}\hat R_k,
\qquad K_\text{fold}=5 .
\end{equation}
The base variance combines the isotonic-calibrated uncertainty from each fold
model with the inter-fold prediction spread,
\begin{equation}
\sigma_{\rm base}^2 =
K_\text{fold}^{-1}\sum_{k=1}^{K_\text{fold}}\sigma_{{\rm cal},k}^2
+ \operatorname{Var}(\hat R_1,\ldots,\hat R_{K_\text{fold}}).
\end{equation}
This base variance is then passed through the test-time corrections described
in Sec.~\ref{sm:sec:uq}: distance scaling, the additive floor, the global scale,
LCO/LCTO inflation, and the tail correction. Distance shrinkage uses
$\gamma=1.0$.

Of the 1008 nuclei, 648 (64\,\%) have $\sigma_{95} < 50$\,mfm, a scale
relevant to present charge-radius measurements at radioactive-ion-beam facilities \cite{sm:Campbell2016,sm:Neugart2017}.
The most certain entries reach $\sigma_{95} \approx 5$--$8$\,mfm in heavier,
well-measured regions of the chart.

The catalogue, the top-twenty table and the catalogue map are built from the
pointwise median of the 45-seed pool rather than from one partition, because
neither the count nor the ranking is determined by a single run. The count
ranges from 513 to 760 across the pool, with a per-seed median of 651; the
reference seed returns 760, the pool maximum. The median catalogue gives 648,
within three of the per-seed median, so pooling introduces no optimism of its
own. The ranking is the more partition-sensitive of the two: per-seed
top-twenty sets agree with one another on a median of 13 of 20 entries (range
6--18), 65 distinct nuclei appear in at least one of them, and only 5 appear in
at least 90\,\% of them. The median ranking overlaps the individual per-seed
rankings on 14.6 of 20 entries on average, against 12.0 for the reference seed,
so it is the more representative summary. A ranking \emph{of} medians is well
defined; a median \emph{of} rankings would not be. Catalogue predictions inherit the regional
reliability of the OOF evaluation: they are most reliable where the OOF
analysis shows smooth residual evolution and nearby measured neighbours, should
be treated more cautiously in the Ca--Ni region where smooth interpolation
reaches its limit, and are indicative rather than fully calibrated for
$Z < 20$, where cluster and continuum effects dominate.

\medskip
\noindent\textbf{Supplementary Data file.} Calibrated ARCUS charge-radius
predictions for the 1008 WS*-based catalogue nuclei defined above are released as
the comma-separated file
\textit{ARCUS\_Charge\_Radii\_Catalogue.csv}, with one
nucleus per row and a header line.
The eight columns are: proton number $Z$, neutron number $N$, and
mass number $A$; the WS* baseline radius $R_\text{baseline}$; the signed ARCUS
residual correction; the calibrated predicted radius $R_\text{ARCUS}$; and the
calibrated Gaussian scale $\sigma_{68}$ together with the $95\%$
prediction-interval half-width $\sigma_{95}$, following the convention of
Eq.~\ref{sm:eq:tau}. All radii and uncertainties are given in fm. The internal kernel
diagnostics used to compute the intervals ($d_\text{aniso}$, $n_\text{eff}$,
and the neutron distance to the measured chain boundary) are omitted from the
released file for clarity.

\end{document}